\documentclass[%
 reprint,
 amsmath,amssymb,
prfluids,
onecolumn
]{revtex4-2}

\usepackage{graphicx}
\usepackage{dcolumn}
\usepackage[hidelinks]{hyperref}

\usepackage[dvipsnames]{xcolor}
\usepackage[capitalise]{cleveref}
\usepackage{bm}
\usepackage{mathrsfs}
\usepackage{subcaption}
\usepackage{todonotes}

\usepackage[normalem]{ulem}

\DeclareMathAlphabet{\mathhv}{OT1}{phv}{m}{it}
\DeclareMathAlphabet{\mathhvb}{OT1}{phv}{bx}{it}

\newcommand{\vect}[1]{\bm{#1}} 
\newcommand{\tens}[1]{\mathhvb{#1}} 
\newcommand{\op}[1]{\mathcal{#1}} 
\newcommand{\tenscomp}[1]{\mathhv{#1}} 
\newcommand{\Rey}{\mathit{Re}_\tau} 
\newcommand{\tr}{\mathrm{tr}} 
\newcommand{\diag}{\mathrm{diag}}
\newcommand{\trsp}{\mathrm{T}} 
\newcommand{\adj}{\dagger} 
\newcommand{\inv}{{-1}} 
\newcommand{\blankop}{(\mathbin{\,\cdot\,})} 
\newcommand{\diff}{\mathrm{d}} 
\newcommand{\Dop}{\mathcal{D}} 
\renewcommand{\Re}{\mathrm{Re}} 
\newcommand{\res}{\mathcal{H}} 
\newcommand{\lap}{\varDelta} 
\newcommand{\lapinv}{\varDelta^{-1}} 
\newcommand{\bnab}{\boldsymbol{\nabla}} 
\newcommand{\OS}{\mathrm{OS}}
\newcommand{\SQ}{\mathrm{SQ}}

\newcommand{\resp}{$\begin{pmatrix}\hat{v} & \hat{\eta}\end{pmatrix}^\trsp$}
\newcommand{\out}[1]{\tilde{#1}} 
\newcommand{\ssim}[1]{\check{#1}} 
\newcommand{\yc}{y_c}
\newcommand{\ycp}{y_c^+}

\begin{document}


\title{
	Interaction of forced Orr-Sommerfeld and Squire modes\\ in a low-order representation of 
	turbulent channel flow 
	}

\author{Ryan M. McMullen}
\email{mcmullen@caltech.edu}
\author{Kevin Rosenberg\footnote[1]{Present address: U.S. Air Force Research Laboratory, Wright–Patterson Air Force Base, OH 45433, USA}}
\author{Beverley J. McKeon}
\affiliation{%
 Graduate Aerospace Laboratories, California Institute of Technology,
 Pasadena, CA 91125, USA}%

\date{\today}

\begin{abstract}
A resolvent-based reduced-order representation is used to capture time-averaged second-order statistics in turbulent channel flow. The recently-proposed decomposition of the resolvent operator into two distinct families related to the Orr-Sommerfeld and Squire operators [K. Rosenberg and B. J. McKeon, Efficient representation of exact coherent states of the Navier-Stokes equations using resolvent analysis, Fluid Dynamics Research 51, 011401 (2019)] results in dramatic improvement in the ability to match all components of the energy spectra and the $uv$ cospectrum. The success of the new representation relies on the ability of the Squire modes to compete with the vorticity generated by Orr-Sommerfeld modes, which is demonstrated by decomposing the statistics into contributions from each family. It is then shown that this competition can be used to infer a phase relationship between the two sets of modes. Additionally, the relative Reynolds number scalings for the two families of resolvent weights are derived for the universal classes of resolvent modes presented by Moarref \emph{et al.} [R. Moarref, A. S. Sharma, J. A. Tropp, and B. J. McKeon, Model-based scaling of the streamwise energy density in
high-Reynolds-number turbulent channels, Journal of Fluid Mechanics 734, 275 (2013)]. 
{These developments can be viewed as a starting point for further modeling efforts to quantify nonlinear interactions in wall-bounded turbulence.}
\end{abstract}

\maketitle

\section{\label{sec:intro}Introduction and background}
Techniques from linear systems theory applied to wall-bounded turbulent shear flows have met with much success. For example, analyses of the Navier-Stokes equations (NSE) linearized about the turbulent mean velocity predict the spanwise length scales associated with the near-wall cycle and large-scale structures in the outer region of the flow from both a transient growth \cite{delAlamo2006,cossu2009,willis2010} and energy amplification of harmonic and stochastic forcing perspective \cite{hwang2010,willis2010}. More recently, the linearized equations have been used to develop linear estimators \cite{illingworth2018,madhusudanan2019,towne2020} and compute impulse responses~\cite{vadarevu2019} that qualitatively reproduce the coherence and self-similarity of large-scale motions. 

As turbulence is an inherently nonlinear phenomenon, a complete model must account for nonlinear interactions. A common approach to incorporate the effects of nonlinearity into linear models is to augment the linearized equations with an eddy viscosity, such that the turbulent mean profile is fixed as an equilibrium solution of the modified mean momentum equation \cite{delAlamo2006,cossu2009,willis2010,hwang2010,illingworth2018,madhusudanan2019,vadarevu2019}. While this approach justifies linearization about the turbulent mean profile, it precludes the study of finite-amplitude fluctuations, since their nonlinear interactions would feed back on and further alter the mean. Instead of using an eddy viscosity, \citet{zare2017} considered colored-in-time stochastic forcing of the linearized NSE in the problem of completing partially-known second-order statistics. Notably, they demonstrated that their approach can be equivalently represented as a low-rank modification of the original equations.

In a different approach to dealing with nonlinearity, the resolvent analysis framework introduced by \citet{mckeon2010} retains the nonlinear term and interprets it as endogenous forcing of the linear dynamics through triadic interactions with the velocity fluctuations at other wavenumber-frequency combinations. This framework eliminates the need to incorporate an eddy viscosity for self-consistency, as no linearization is performed. 
{\citet{landahl1967} arrived at a similar formulation, deriving a forced Orr-Sommerfeld equation in the study of wall-pressure fluctuations, but focused on obtaining approximate solutions in eigenfunction expansions.} 

Closure of the loop requires determination of the forcing such that it yields the correct velocity Fourier modes, as well as the mean velocity profile, which is assumed known. The forcing can be expanded as a sum over a set of basis functions such that the unknowns are the complex amplitudes, called the resolvent weights. An exact equation for the weights can be formulated \cite{mckeon2013}, though it is intractable to solve for most complex flows of interest. Consequently, there have been previous attempts to estimate the weights from data, e.g. by using either a single time series or power spectral density of the velocity fluctuations \cite{gomez2016,beneddine2016}. Alternatively, \citet{moarref2014} used convex optimization to compute the weights for a resolvent-based low-order representation of time-averaged velocity spectra that minimize the deviation from spectra obtained from a direct numerical simulation (DNS) of $\Rey=2003$ channel flow \cite{hoyas2006}. \citet{towne2018} established a link between resolvent analysis and spectral proper orthogonal decomposition (SPOD) and showed that if the resolvent weights are treated as stochastic quantities, their covariance matrix can be calculated from the SPOD modes, which inherently rely on statistical data.

{In special cases where full information of the nonlinear forcing is available, such as for exact coherent states (ECS), the resolvent weights can be computed exactly by projecting the forcing onto the aforementioned set of basis functions \cite{sharma2016}. For ECS families in channel and pipe flow, which come in pairs of upper and lower branch solutions, the lower branch ones are typically well-represented by only a few resolvent modes, whereas many of the upper branch solutions are not captured as efficiently. Furthermore, the wall-normal and spanwise velocity components converge much more slowly than the streamwise velocity. However, an alternative decomposition of the resolvent operator recently proposed  by \citet{rosenberg2019a} yields two families of modes related to the Orr-Sommerfeld and Squire operators from classical linear stability theory. By projecting the same channel ECS, they demonstrated that the new sets of basis functions enable a much more compact representation of both branches of solutions, and, notably, all three velocity components converge at roughly the same rate. Subsequent analysis attributed the improved efficacy of the alternative decomposition to the isolation of the wall-normal velocity response into the Orr-Sommerfeld modes, such that the Squire wall-normal vorticity is free to interact with that generated by the Orr-Sommerfeld modes \cite{rosenberg2018}.}

{While the utility of the decomposition into Orr-Sommerfeld and Squire modes for highly simplified flows like ECS has been established, an open question is whether or not it remains relevant for high Reynolds number turbulence. In the present work, it is shown that the second-order statistics of turbulent channel flow can be accurately represented using a low-order approximation based on this framework. It is additionally shown that the vorticity produced by the Orr-Sommerfeld and Squire modes act to oppose each other, and this observation reveals information about how the resolvent weights for the two families scale relative to each other with Reynolds number. Altogether, these insights point to a mechanism in turbulent channel flow that is important for low-order modeling efforts.}

\section{\label{sec:form}Formulation}
\subsection{\label{sec:res}Resolvent analysis of turbulent channel flow}
The approach is based on the resolvent analysis framework of \citet{mckeon2010}, in which the incompressible NSE,
\begin{subequations} 
\label{eq:NSE}
\begin{align}
\partial_t \tilde{\vect{u}} + \left( \tilde{\vect{u}} \boldsymbol{\cdot} \bnab \right) \tilde{\vect{u}} = - \bnab \tilde{p} + \Rey^\inv\bnab^2 \tilde{\vect{u}}&,
\\
\bnab \boldsymbol{\cdot} \tilde{\vect{u}} = 0&,
\end{align}
\end{subequations}
here nondimensionalized using the friction velocity $u_\tau$ and channel half-height $h$, are first Reynolds decomposed as $\tilde{\vect{u}} = \vect{U} + \vect{u}$, where $\vect{U} = \begin{pmatrix}U(y) & 0 & 0\end{pmatrix}^\trsp$ is the turbulent mean velocity profile and $\vect{u}$ are the fluctuations about the mean, and then Fourier transformed in the homogeneous wall-parallel and temporal directions $x$, $z$, and $t$, resulting in equations for the Fourier coefficients, denoted by $\hat{\blankop}\,$. For each wavenumber-frequency triplet $\begin{pmatrix}k_x & k_z & \omega\end{pmatrix}^\trsp \neq \mathbf{0}$, we have
\begin{subequations}
\begin{align}
i\omega \hat{\vect{u}} + \left( \vect{U} \boldsymbol{\cdot} \bnab \right) \hat{\vect{u}} + \left( \hat{\vect{u}} \boldsymbol{\cdot} \bnab \right) \vect{U} + \bnab \hat{p} - \Rey^\inv\bnab^2 \hat{\vect{u}} =& \hat{\vect{f}}
\\
\bnab \boldsymbol{\cdot} \hat{\vect{u}} =& 0,
\end{align}
\label{eq:NSEf}
\end{subequations}
where $\vect{f} = -\left( \vect{u} \boldsymbol{\cdot} \bnab \right) \vect{u} + \langle \left( \vect{u} \boldsymbol{\cdot} \bnab \right) \vect{u} \rangle$ and $\langle\,\cdot\,\rangle$ denotes an averaged quantity, is interpreted as a forcing that drives the dynamics linear in $\hat{\vect{u}}$. The pressure can be projected out of \cref{eq:NSEf} using the standard mapping to wall-normal velocity $\hat{v}$ and wall-normal vorticity $\hat{\eta} = ik_z\hat{u} - ik_x \hat{w}$. The equations are then concisely written as
\begin{equation}
    \begin{pmatrix}
    \hat{v}\\
    \hat{\eta}
    \end{pmatrix}
    = \res\!\left( k_x,\, k_z,\, \omega \right) 
    \hat{\vect{g}},
\label{eq:NSEres}
\end{equation}
where 
\begin{equation}
    \mathcal{H} = 
    \begin{pmatrix}
    -i \omega - \lapinv \mathcal{L}^\OS & 0\\
    -i k_z U' & -i \omega - \mathcal{L}^\SQ
    \end{pmatrix}^{-1}
    \label{eq:resop}
\end{equation}
is the resolvent operator, $\lap = \Dop^2-k^2$, $\Dop=\diff/\diff y$, $k^2=k_x^2+k_z^2$, and $U'=\Dop U$. Additionally,
\begin{subequations}
\label{eq:Lossq}
	\begin{align}
	    \mathcal{L}^\OS &= i k_x\!\left(U'' - U\lap\right) + \Rey^\inv\lap^2,
	    \label{eq:Los}\\
	    \mathcal{L}^\SQ &= -i k_x U + \Rey^\inv\lap
	    \label{eq:Lsq}
	\end{align}
\end{subequations}
are the Orr-Sommerfeld (OS) and Squire (SQ) operators, respectively. The forcing term $\hat{\vect{g}} = \begin{pmatrix} \hat{g}_v & \hat{g}_\eta \end{pmatrix}^\trsp$ in \cref{eq:NSEres} is related to $\hat{\vect{f}}$ via
\begin{equation}
    \hat{\vect{g}}
    =
    \underbrace{
    \begin{pmatrix}
    -i k_x \lapinv \Dop  & -k^2\lapinv & -i k_z \lapinv \Dop\\
    i k_z & 0 & -i k_x
    \end{pmatrix}
	}_{\textstyle \mathcal{B}}
    \hat{\vect{f}}.
    \label{eq:forcing}
\end{equation}
Note that $\hat{\vect{g}}$ is solenoidal, since the irrotational component of $\hat{\vect{f}}$ lies in the null space of $\mathcal{B}$ \cite{rosenberg2019a}. 

With the aim of obtaining a low-order representation  of \cref{eq:NSEres}, we compute the Schmidt decomposition of $\res$:
\begin{equation}
\res = \sum_{j=1}^\infty \vect{\psi}_j \sigma_j \langle \,\cdot\,,\vect{\phi}_j\rangle,
\label{eq:svd}
\end{equation}
where $\sigma_j \geq \sigma_{j+1}\geq 0$ $\forall j$ are the singular values, and $\vect{\psi}_j$ and $\vect{\phi}_j$ are the left and right singular vectors, respectively. Since the $\vect{\psi}_j$ are a basis for the output space, i.e., the space to which the response \resp belongs, they are referred to as response modes; similarly, the $\vect{\phi}_j$ are referred to as the forcing modes. The Schmidt decomposition applies to linear operators on infinite-dimensional vector spaces. For the finite-dimensional matrix approximation obtained from numerically discretizing the operator, this becomes the singular value decomposition (SVD), which we refer to hereafter for simplicity. As is evident from \cref{eq:svd}, the SVD depends on the choice of inner product $\langle \,\cdot\, , \,\cdot\, \rangle$. On both the input and output spaces we adopt the standard kinetic energy inner product \cite{schmid2001}:
\begin{equation}
\langle \vect{x}_1, \vect{x}_2 \rangle = \int_{-1}^1 \vect{x}_2^* \mathcal{Q} \vect{x}_1 \diff y,
\label{eq:IP}
\end{equation}
where $\blankop^*$ denotes the conjugate transpose and 
\begin{equation}
\mathcal{Q} = \frac{1}{k^2}
\begin{pmatrix}
-\lap & 0\\
0 & 1
\end{pmatrix}.
\label{eq:Q}
\end{equation}
The left and right singular vectors are orthonormal with respect to this inner product, i.e., $\langle \vect{\psi}_j, \vect{\psi}_k \rangle = \langle \vect{\phi}_j, \vect{\phi}_k \rangle = \delta_{jk}$, where $\delta_{jk}$ is the Kronecker delta. 

The desired low-order approximation of \cref{eq:NSEres} is obtained by truncating the sum in \cref{eq:svd}:
\begin{equation}
\begin{pmatrix}
\hat{v}\\
\hat{\eta}
\end{pmatrix} 
\approx \sum_{j=1}^N \vect{\psi}_j \sigma_j  \chi_j,
\label{eq:approx}
\end{equation}
for some $N\geq 1$. We refer to this as the rank-$N$ approximation. The $\chi_j = \langle \hat{\vect{g}},\vect{\phi}_j\rangle$ are called the resolvent weights and quantify how much of the forcing $\hat{\vect{g}}$ is in the direction $\vect{\phi}_j$. {For broadband forcing in $y$, i.e., $\chi_j = \chi$ $\forall j$, \cref{eq:approx} is optimal in the norm induced by the inner product in \cref{eq:IP}. 
Furthermore, if $\sum_{j=1}^N \sigma_j^2 \approx \sum_{j=1}^\infty \sigma_j^2$ for relatively small $N$, 
$\res$ is said to be effectively low-rank.} 
It has been shown that this property holds for a large portion of spectral space that is energetically significant \cite{moarref2013}, and this low-rank behavior has previously been exploited to model salient features in wall-bounded turbulence \cite{mckeon2010,sharma2013,moarref2013}. {However, the assumption of broadband forcing is in general not valid. For the case of structured forcing, \cref{eq:approx} is an accurate approximation of the full system, provided the forcing is not too aligned in any of the truncated directions.}

Finally, in order to compute the second-order velocity statistics, the velocity $\hat{\vect{u}}$ is recovered from the response via
\begin{equation}
    \hat{\vect{u}}
    =
    \frac{1}{k^2}
    \begin{pmatrix}
    i k_x \Dop & -i k_z\\
    k^2 & 0\\
    i k_z \Dop & i k_x
    \end{pmatrix}
    \begin{pmatrix}
    \hat{v}\\
    \hat{\eta}
    \end{pmatrix}.
    \label{eq:uvw}
\end{equation}

\subsection{\label{sec:OSSQ}Orr-Sommerfeld and Squire decomposition of the resolvent}
As discussed in \cref{sec:res}, the decomposition of $\res$ given in \cref{eq:svd}, hereafter referred to as the standard resolvent decomposition, is optimal in the kinetic energy norm induced by the inner product in \cref{eq:IP}. However, in wall-bounded turbulence the kinetic energy is often dominated by the the streamwise velocity, which means that all three velocity components may not be approximated uniformly well \cite{moarref2014,sharma2016}. In such situations, an alternative decomposition that more faithfully represents the underlying dynamics may be desirable. This idea has been explored previously by \citet{juttijudata2005}, who transformed near-wall data from turbulent channel flow into Squire's coordinate system and then performed POD on modes associated with the streamwise streaks and rolls separately. While the resulting basis functions are energetically suboptimal compared to those from standard POD, they demonstrate that the reconstruction of wall-normal, spanwise, and Reynolds shear stress statistics improve substantially. 

In a similar spirit, \citet{rosenberg2019a}, proposed the following alternative decomposition of $\res$. Note that \cref{eq:NSEres} can be rewritten as
\begin{equation}
	\begin{pmatrix}
	\hat{v}\\
	\hat{\eta}
	\end{pmatrix}
	=
	\begin{pmatrix}
	\res_{vv} & 0\\
	\res_{\eta v} & \res_{\eta\eta}
	\end{pmatrix}
	\begin{pmatrix}
	\hat{g}_v\\
	\hat{g}_\eta
	\end{pmatrix},
	\label{eq:NSEOS}	 
\end{equation}
where
\begin{subequations}
\label{eq:Hops}
	\begin{align}
	\res_{vv} &= \left( -i\omega - \lapinv \mathcal{L}^\OS \right)^{-1},
	\label{eq:Hvv}\\
	\res_{\eta\eta} &= \left( -i\omega - \mathcal{L}^\SQ \right)^{-1},
	\label{eq:Hetaeta}\\
	\res_{\eta v} &= -i k_z \res_{\eta \eta} U' \res_{vv}.
	\label{eq:Hetav}
	\end{align}
\end{subequations}	
Apparently, $\res_{vv}$ and $\res_{\eta v}$ are forced by $\hat{g}_v$ only, while $\res_{\eta \eta}$ is forced by $\hat{g}_\eta$ only. This motivates the separation of the response \resp into two distinct families:
\begin{subequations}
\label{eq:OSSQ}	
	\begin{align}
		\begin{pmatrix}
		\hat{v}\\
		\hat{\eta}^\OS
		\end{pmatrix}
		&=
		\begin{pmatrix}
		\res_{vv}\\
		\res_{\eta v}
		\end{pmatrix}
		\hat{g}_v,
		\label{eq:resOS}\\
		\hat{\eta}^\SQ &= \res_{\eta\eta}\,
		\hat{g}_\eta.
		\label{eq:resSQ}
	\end{align}
\end{subequations}
In the following, we refer to the family of modes in \cref{eq:resOS} as Orr-Sommerfeld (OS) modes and the family in \cref{eq:resSQ} as Squire (SQ) modes. {The separation of $\hat{\eta}$ into two distinct families is common practice in linear stability analysis, where the SQ and OS modes are, respectively, the homogeneous and particular solutions of the Squire equation: $(-i\omega - \op{L}^\SQ)\hat{\eta}=-ik_zU'\hat{v}$ \cite{schmid2001}. That is, the OS modes can be interpreted as a response to the wall-normal velocity. This interpretation still holds in the nonlinear setting, since the second component of \cref{eq:resOS} can be written as $\hat{\eta}^\OS = -ik_z\res_{\eta \eta}U'\hat{v}$. However, the SQ modes are no longer the homogeneous solutions, but are now interpreted as the response to forcing by $\hat{g}_\eta$.}

{Note that only the OS modes contain a $\hat{v}$ response, such that the SQ modes contribute only to the $\hat{\eta}$ response, i.e., to the wall-parallel velocity components. There is thus the potential for interaction between the OS and SQ vorticity in ways that are not admitted by the standard resolvent decomposition. This fact is of central importance for the OS-SQ resolvent decomposition, and it will be demonstrated in \cref{sec:kxspectra} that this drastically improves the accuracy of a low-order resolvent-based representation of the second-order statistics for turbulent channel flow.} 

An SVD of each operator in \cref{eq:OSSQ} is performed separately, and the resulting decomposition is referred to as the OS-SQ decomposition of the resolvent. The approximation of the response becomes
\begin{equation}
\begin{pmatrix}
\hat{v}\\
\hat{\eta}
\end{pmatrix} 
\approx \sum_{j=1}^{N^\OS} \vect{\psi}^\OS_j \sigma^\OS_j  \chi^\OS_j + \sum_{k=1}^{N^\SQ} \vect{\psi}^\SQ_k \sigma^\SQ_k  \chi^\SQ_k.
\label{eq:seriesOSSQ}
\end{equation}
Note that \cref{eq:seriesOSSQ} is now a sum of $N^\OS + N^\SQ$ terms. Furthermore, while the left and right singular vectors of each family still comprise orthonormal sets with respect to the inner product given in \cref{eq:IP}, it is not guaranteed that modes belonging to different families are orthonormal, e.g. $\langle \vect{\psi}^\OS_j, \vect{\psi}^\SQ_k \rangle \neq \delta_{jk}$ in general. 

\subsection{\label{sec:opt}Emprical determination of the resolvent weights via convex optimization}
The singular values and vectors are computed directly from the resolvent operator, which depends only on the (assumed known) mean velocity profile $U$, whereas computation of the weights requires solution of a nonlinear programming problem 
\cite{mckeon2013}. This can be done exactly in special cases, such as for exact coherent states (ECS) \cite{rosenberg2018}. However, it rapidly becomes intractable with an increasing number of degrees of freedom, and, to our knowledge, fully turbulent flows remain out of reach. 

Consequently, several attempts have been made to determine the weights empirically \cite{moarref2014,gomez2016,beneddine2016,zare2017,towne2018}. In particular, \citet{moarref2014}, used convex optimization to compute the weights that minimize the deviation between a resolvent-based representation of the energy spectra and DNS data. We take the same approach here and largely adopt their formulation, with the major exception that we employ the OS-SQ decomposition discussed in \cref{sec:OSSQ}. That is, for given $N^\OS$ and $N^\SQ$, we attempt to approximate the DNS statistics using the approximation given in \cref{eq:seriesOSSQ}.

As introduced by \citet{moarref2014}, the resolvent three-dimensional streamwise energy spectra are
\begin{equation}
E_r(y, k_x, k_z, c) = \Re\!\left\{ \tr\! \left( \tens{A}_r\tens{X} \right) \right\},
\label{eq:E}
\end{equation}
with $r\in\{uu, vv, ww, uv\}$, and where $\Re\{\,\cdot\,\}$ is the real part of a complex number and $\tr\blankop$ is the matrix trace. Note that we have chosen to parameterize the spectra in terms of the wavespeed $c=\omega/k_x$ since resolvent modes tend to be localized about the critical layers $y_c$, where $U(y_c) = c$ \cite{mckeon2010}, and it has been observed experimentally that the range of energetic wavespeeds is relatively compact, with the most energetic motions typically being confined to the range $8 \lesssim c \lesssim U_{cl}$ \cite{lehew2011}, where $U_{cl}$ is the mean centerline velocity. In \cref{eq:E}, the matrix $\tens{A}_{uu}$, for example, with entries
\begin{equation}
\tenscomp{A}_{uu,ij} = \sigma_i\sigma_j \hat{u}_i \hat{u}_j^*,
\label{eq:A}
\end{equation}
represents the contributions of the singular values and response modes and can be determined \emph{a priori} from the SVD of the resolvent. The matrix $\tens{X}$, with entries
\begin{equation}
\tenscomp{X}_{ij} = \chi_i^*\chi_j,
\label{eq:X}
\end{equation}
is the weights matrix. Apparent from this definition is that $\tens{X}^{\,\trsp} = \vect{\chi}\vect{\chi}^* \succeq \vect{0}$, where $\vect{\chi}$ is the vector of weights and $\succeq$ denotes the L\"{o}wner order, i.e., $\tens{X}$ is a rank-1 positive-semidefinite matrix. 
The OS-SQ decomposition is incorporated into this framework simply by partitioning the $\tens{A}_r$ and $\tens{X}$ matrices as
\begin{align}
&\tens{A}_r = 
\begin{pmatrix}
\tens{A}_r^{\OS/\OS} & \tens{A}_r^{\OS/\SQ}\\
\tens{A}_r^{\SQ/\OS} & \tens{A}_r^{\SQ/\SQ}
\end{pmatrix},
&\tens{X} = 
\begin{pmatrix}
\tens{X}^{\OS/\OS} & \tens{X}^{\OS/\SQ}\\
\tens{X}^{\OS/\SQ\,*} & \tens{X}^{\SQ/\SQ}
\end{pmatrix},
\label{eq:partition}
\end{align}
where the superscript $\text{X}/\text{Y}$ denotes the family of the $i$th and $j$th mode, respectively, in \cref{eq:A,eq:X}. 

The goal is to compute the weights matrix such that the deviation between the wavespeed-integrated resolvent spectra in \cref{eq:E} and time-averaged DNS spectra is minimized. After discretization of the wavespeed range $c\in [0,U_{cl}]$, this can be formally cast as the following optimization problem: For fixed $k_x$ and $k_z$,
\begin{equation}
\begin{aligned}
&\underset{\{\tens{X}_l\}_{l = 1,2,\dots,N_c}, \, e}{\text{minimize}} &&e
\\
&\text{subject to} &&\frac{\lVert E^{\text{DNS}}_r - \sum_{l=1}^{N_c} k_x\,\diff c\, \Re\!\left\{ \tr\!\left( \tens{A}_{r,l}\tens{X}_l\right) \right\} \rVert^2}{\lVert E^{\text{DNS}}_r \rVert^2} \leq e
\label{eq:optprob}\\
&&&\tens{X}_l \succeq \vect{0},\, l = 1,2,\dots,N_c,
\end{aligned} 
\end{equation}
where the subscript $l$ denotes a quantity evaluated at $c=c_l$. Note that the norm $\lVert \,\cdot\, \rVert$ is not the one induced by \cref{eq:IP}. It is defined as
\begin{equation}
\left\Vert f \right\Vert^2 = \int_{y_{\min}^+}^{y_{\max}^+} \left\vert f(\log y^+) \right\vert^2 \diff \log y^+
\end{equation}
and is designed to penalize deviations across the channel equally \cite{moarref2014}. Thus deviations from the DNS spectra are enforced for $5\leq y_{\min}^+ \leq y^+ \leq y_{\max}^+<\Rey$.

\cref{eq:optprob} is a semidefinite program for the weights matrices $\tens{X}_l$ and can therefore be solved efficiently using a convex optimization software package. Note that imposing the rank-1 constraint on the $\tens{X}_l$ would make \cref{eq:optprob} non-convex. \citet{moarref2014} employed an iterative rank-reduction procedure to recover rank-1 matrices from the full-rank solution \cite{huang2009}. However, we do not employ this algorithm here and instead choose to work with the full-rank weights matrices. In this case, the $\tens{X}_l$ can be interpreted as the covariance matrices of the weights, similar to \citet{towne2018}. {Finally, since the optimization is performed for second-order statistics, the present approach does not provide phase information about modes with different wavenumbers. This means that the computed weights do not yield a closed, self-consistent system, as such information is necessary to recover the mean velocity profile {as well as the fluctuations}. Extension of the method to incorporate phase is a direction for future work.}

\subsection{\label{sec:numer}Numerical details}
The resolvent operators are discretized in MATLAB using a Chebyshev pseudospectral method \cite{weideman2000}; all results presented here use 203 Chebyshev polynomials. The SVDs of the discretized operators are performed with a random matrix algorithm, which is faster than MATLAB's built-in \texttt{svd()} \cite{halko2011}. The time-averaged two-dimensional DNS spectra for $\Rey=934$ and $\Rey=2003$ are obtained from \citet{hoyas2006}, and for $\Rey=4219$ from \citet{lozano2014}. Additionally, spectra were generated for $\Rey=185$ using Channelflow \cite{channelflow}. The resolutions of all the DNS considered are given in \cref{tab:DNS}. 

For the results presented in \cref{sec:kxspectra}, which focus on $\Rey=2003$, the DNS spectra are interpolated onto a grid of $N_{k_x}=30$ by $N_{k_z}=31$ logarithmically spaced wavenumbers, {which is sufficient to reproduce the general shape of the spectra. Furthermore, it has recently been shown that statistics such as the $uv$ Reynolds stress can be accurately reproduced even when retaining only about 2\% of the wavenumbers from DNS \cite{toedtli2019}}. Both the spectra and resolvent modes are interpolated onto a common grid of $N_y=60$ logarithmically spaced points in the wall-normal direction, and the wavespeed range $c \in [0, U_{cl}]$ is discretized into $N_c=100$ linearly spaced wavespeeds. The optimization problem \cref{eq:optprob} is then solved with CVX \cite{cvx}. The results are insensitive to further increases in $N_y$ and $N_c$ \cite{moarref2014}.

\begin{table}[h]
	\caption{\label{tab:DNS}Resolutions of the DNS from which spectra were obtained for the optimization.}
	\begin{ruledtabular}
		\begin{tabular}{lrrr}
			\noalign{\smallskip}
			$\Rey$ & $N_x$ & $N_y$ & $N_z$\\
			\noalign{\medskip}
			185 & 384 & 129 & 128\\
			934 \cite{hoyas2006} & 1024 & 385 & 768\\
			2003 \cite{hoyas2006} & 2048 & 635 & 1536\\
			4219 \cite{lozano2014} & 1024 & 1081 & 1024\\
			\noalign{\smallskip}
		\end{tabular}
	\end{ruledtabular}
\end{table}

\section{Analysis of the optimized spectra}

\subsection{\label{sec:kxspectra}Reconstruction of time-averaged statistics}
\begin{figure}
    \centering \includegraphics[width=0.62\textwidth]{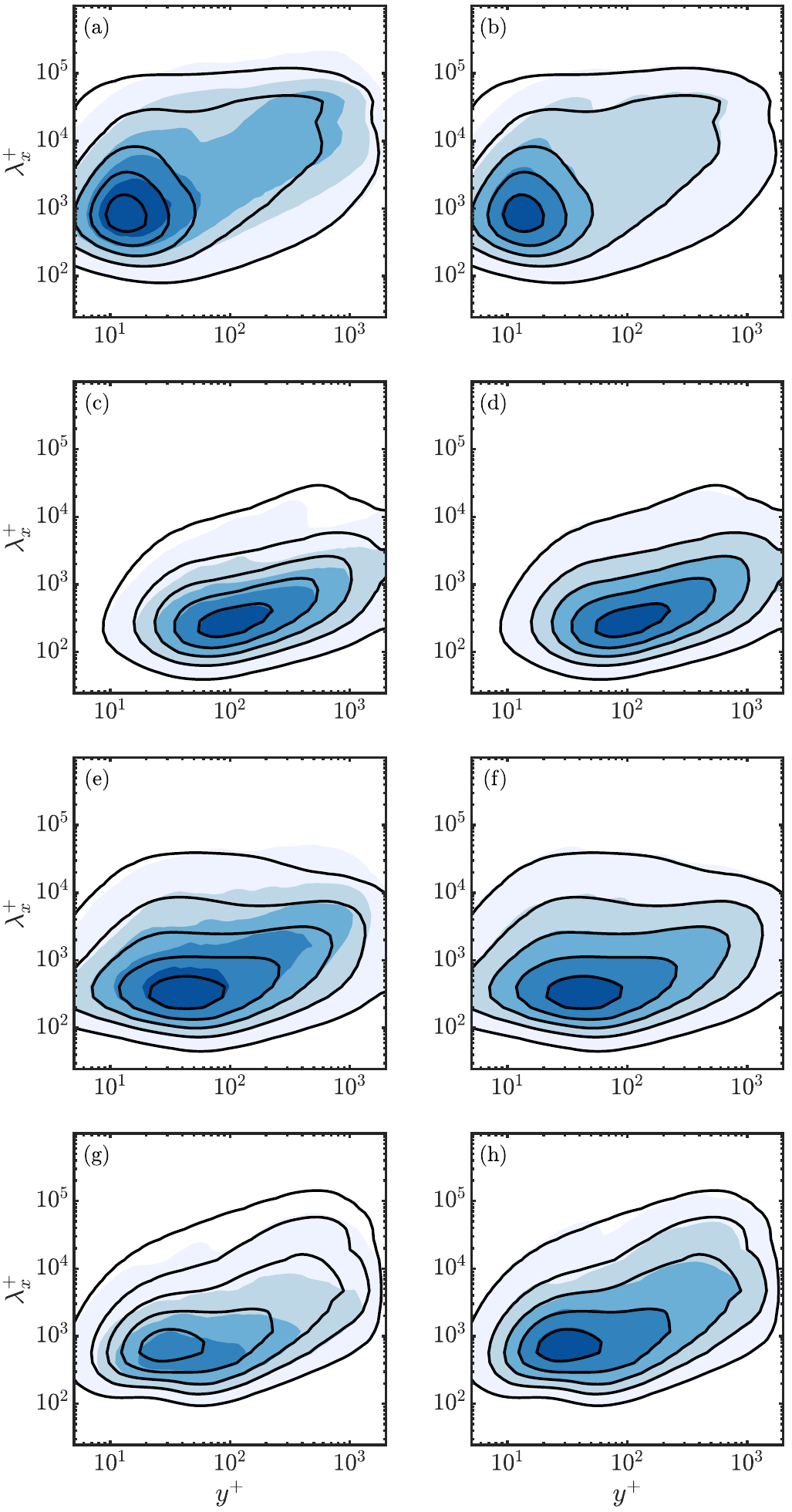}
    \caption{Premultiplied one-dimensional spectra from the resolvent (filled contours) and DNS (black contours) for $\Rey=2003$. (a,c,e,g) Standard resolvent decomposition using $N=6$ modes per wavenumber-frequency triplet; (b,d,f,h) OS-SQ resolvent decomposition using $N^{\OS} = N^{\SQ} = 3$ modes per wavenumber-frequency triplet. (a,b) $k_xE_{uu}$, (c,d) $k_xE_{vv}$, (e,f) $k_xE_{ww}$, (g,h) $-k_xE_{uv}$. Contour levels are from 10\% to 90\% of the DNS maximum in 20\% increments.\label{fig:kxspect}
    }
\end{figure}

\begin{figure}[h]
    \centering \includegraphics[width=0.65\textwidth]{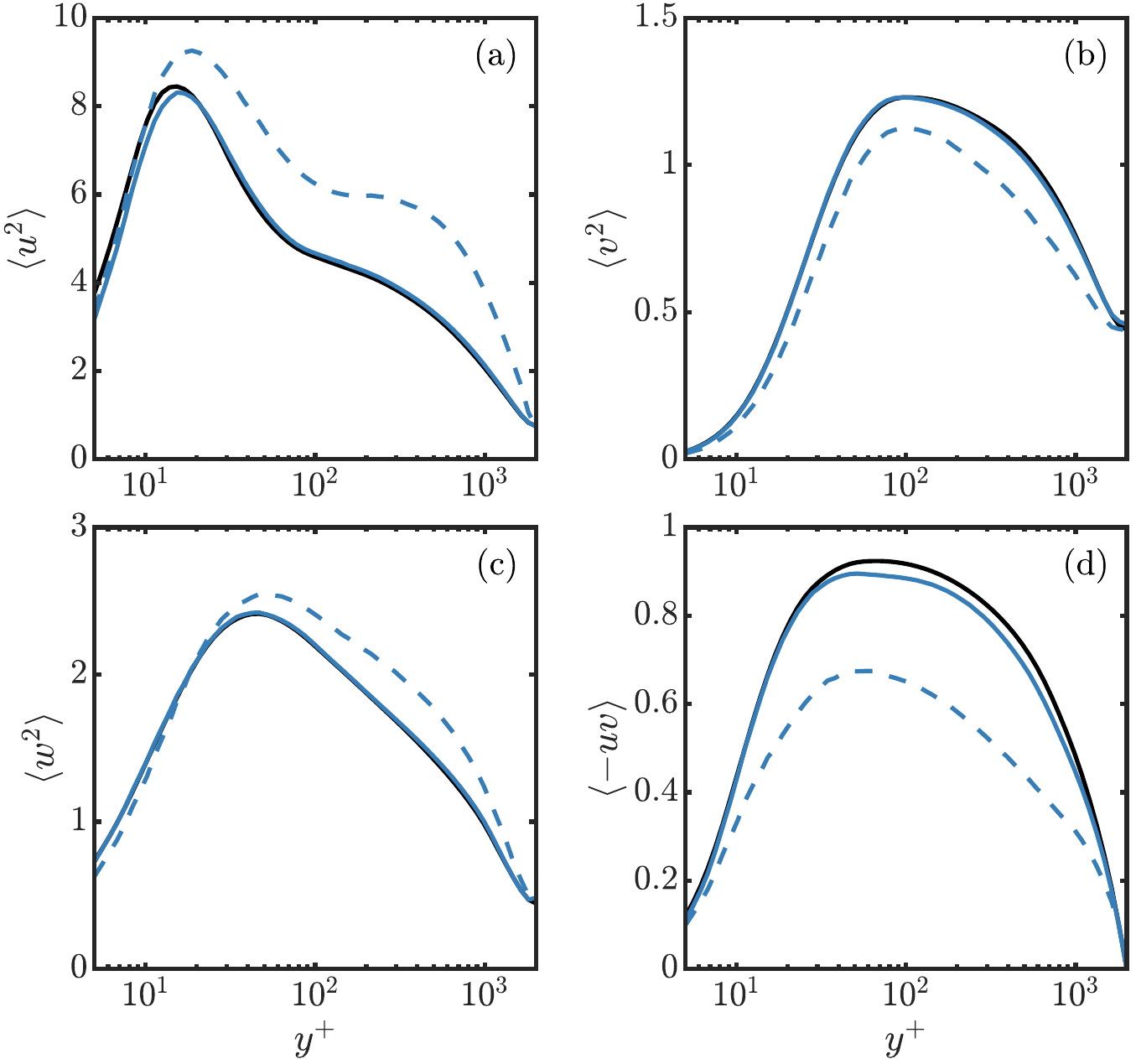}
    \caption{Intensities from the resolvent with $N^{\OS}=N^{\SQ}=3$ modes per wavenumber-frequency triplet (blue) and DNS (black) for $\Rey=2003$. Also shown in dashed lines are the intensities obtained from the standard resolvent decomposition approach using the same total number of modes.\label{fig:ints}}
\end{figure}

The accuracy of the optimized spectra is evaluated by comparing them to the time-averaged statistics from the DNS for $\Rey=2003$. The premultiplied 1D $k_x$ spectra, 
\begin{equation}
k_xE_r(y,\, k_x) = \int_{k_{z,\mathrm{min}}}^{k_{z,\mathrm{max}}} \int_{0}^{U_{\mathit{cl}}} k_x^2 \, E_r(y, k_x, k_z, c) \,\diff c \, \diff k_z, 
\end{equation}
using $N^{\OS}=N^{\SQ}=3$ modes, i.e., six modes per wavenumber-frequency triplet, are compared to the DNS in the right column of \cref{fig:kxspect}, which is plotted in terms of $\lambda_x^+ = 2\pi/k_x^+$. Clearly, $N^{\OS}=N^{\SQ}=3$ modes is sufficient to accurately reproduce the spectra since the overall agreement between the resolvent and DNS spectra is very good, and in particular, the peaks are captured almost exactly. The only significant discrepancies are in $k_xE_{uu}$ at large $\lambda_x^+$ and $y^+ \lesssim 100$ and $-k_xE_{uv}$ at large $\lambda_x^+$ and $y^+ \lesssim 50$. Further discussion of these discrepancies, as well as the accuracy of the optimized spectra using different numbers of modes is given in \cref{sec:err}. Also shown in the left column are the spectra obtained using the standard decomposition with the same total number of modes. The performance is significantly worse, with $k_xE_{uu}$ and $k_xE_{ww}$ being greatly over-predicted, and $k_xE_{vv}$ and $-k_xE_{uv}$ being under-predicted. In fact, the standard resolvent decomposition fails to capture the 90\% energy level (darkest blue contours) for $-k_xE_{uv}$. 
Subsequent integration over $k_x$ gives the intensities, which are shown in figure \cref{fig:ints}. The deviation errors are 4.3\%, 0.95\%, 0.66\%, and 3.8\% for $\langle u^2\rangle$, $\langle v^2\rangle$, $\langle w^2\rangle$, and $\langle -uv \rangle$, respectively. These should be compared with errors of 30\%, 14\%, 12\%, and 31\% using the standard resolvent decomposition, shown in the dashed curves. 

{As the goal of the optimized spectra is to obtain a \emph{low-order} representation of the spectra, it is worth comparing the number of degrees of freedom of the resolvent spectra to the original DNS. For a given $k_x,k_z$, the $\Rey=2003$ DNS spectra were computed using $N_y=665$ wall-normal grid points and $N_t=7730$ snapshots \cite{hoyas2006}. For the results shown in \cref{fig:kxspect,fig:ints}, the resolvent representation was computed using $N=6$ resolvent modes and $N_c=100$ wavespeeds, about 0.01\% of the degrees of freedom in the DNS.}

\subsection{\label{sec:pwrspect}Prediction of additional statistics} 

\begin{figure}
	\centering
	\includegraphics[width=0.7\linewidth]{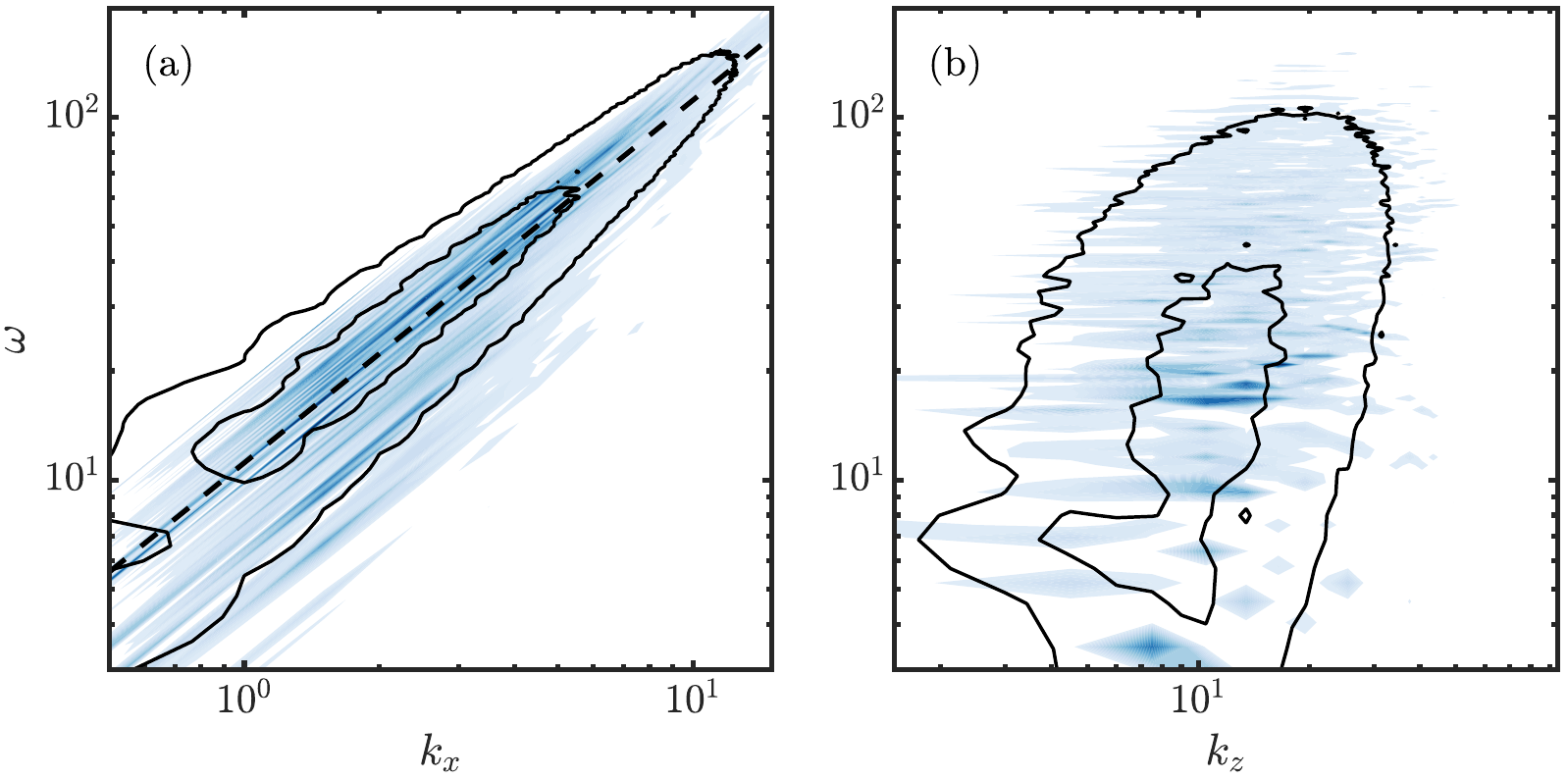}
	\caption{2D premultiplied streamwise velocity power spectra at $y^+\approx15$ for $\Rey=185$. (a) $\omega k_x E_{uu}$, (b) $\omega k_z E_{uu}$. Filled contours: Optimized weights with $N^\OS=N^\SQ=3$. Line contours: DNS; levels are 10\% and 50\% of the maximum value. The slope of the dashed line indicates the the local mean velocity $U(y^+=15)$. 
		\label{fig:pwrspect}}
\end{figure}

Since the optimization only attempts to match time-averaged spectra, the distribution of energetic content in $c$ is not directly constrained. To assess this, the power spectra computed from the resolvent representation using the optimized weights with $N^\OS=N^\SQ=3$ are compared to those computed from DNS for $\Rey=185$ using Welch's method with 8491 snapshots divided into 10 segments having 50\% overlap. \cref{fig:pwrspect}(a) and \cref{fig:pwrspect}(b) show the 2D premultiplied streamwise power spectra in the $k_x-\omega$ and $k_z-\omega$ planes, respectively, at $y^+\approx15$. 

The distribution in the $k_x-\omega$ plane is fairly good, with most of the energetic content of the resolvent spectrum falling within the 10\% DNS contour. As discussed above, the distribution in $c$ is not directly constrained. However, the localization of the leading resolvent modes at the critical layer implies that the energetic content at a given wall-normal location is largely contributed by modes with a wavespeed matching the local mean velocity. This is evident in \cref{fig:pwrspect}, where both the DNS and resolvent spectra are concentrated around the dashed line representing a constant wavespeed $c=U(y^+=15)$. There is no such localization mechanism in the $k_z-\omega$ plane. 
Nonetheless, the resolvent representation still reproduces the general shape of the DNS spectrum quite well. Note that to produce \cref{fig:pwrspect}(b), the resolvent spectrum was interpolated onto a common $\omega$ grid prior to integration over $k_x$.

\begin{figure}
	\includegraphics[width=0.7\linewidth]{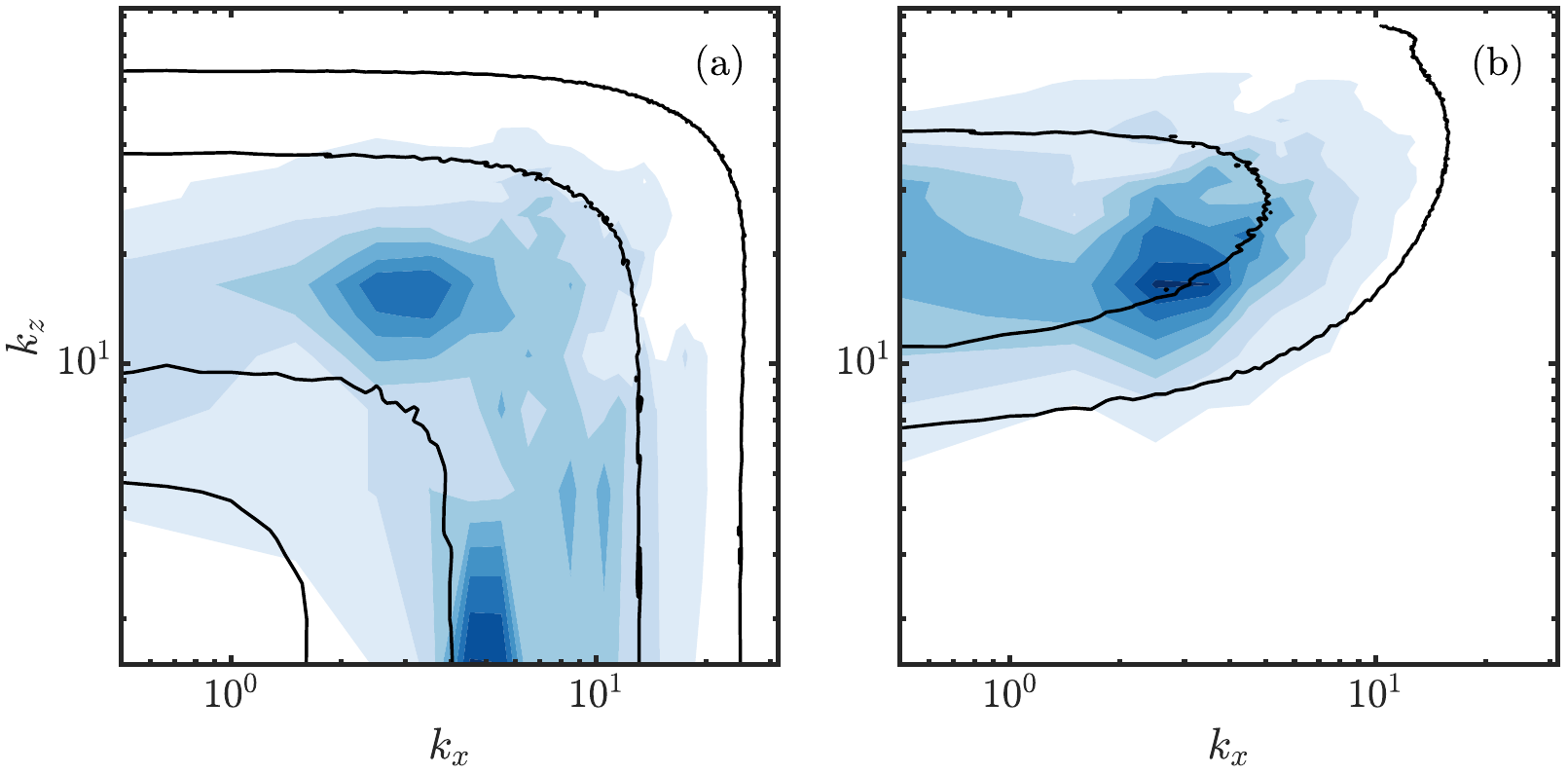}
	\caption{2D forcing spectra at $y^+\approx15$ for $\Rey=185$. (a) $E_{g_v g_v}$, (b) $E_{g_\eta g_\eta}$. Filled contours: Optimized weights with $N^\OS=N^\SQ=3$. Line contours, reproduced from \citet{rosenberg2018}: DNS; levels are 10\% and 50\% of the maximum value. \label{fig:frcspect}}
\end{figure}

The optimized weights can also be used to compute an approximation of the forcing spectra in a manner that is directly analogous to the velocity spectra in \cref{eq:E}:
\begin{subequations}
	\label{eq:Efrc}	
	\begin{align}
	E_{g_vg_v}(y, k_x, k_z, c) &= \Re\!\left\{ \tr\! \left( \tens{B}_{vv}\tens{X}^{\OS/\OS} \right) \right\},
	\\
	E_{g_\eta g_\eta}(y, k_x, k_z, c) &= \Re\!\left\{ \tr\! \left( \tens{B}_{\eta\eta}\tens{X}^{\SQ/\SQ} \right) \right\},
	\end{align}
\end{subequations}
where $\tenscomp{B}_{vv,ij} = \phi_{v,i} \phi_{v,j}^*$, and $\tenscomp{B}_{\eta\eta,ij} = \phi_{\eta,i} \phi_{\eta,j}^*$. The estimates of the 2D forcing spectra with $N^\OS=N^\SQ=3$ in the $k_x - k_z$ plane at $y^+\approx15$ for $\Rey=185$ are compared to the full forcing spectra computed from DNS \cite{rosenberg2018} in \cref{fig:frcspect}. The resolvent estimate reasonably predicts the general shape of the full spectra with only a few modes; this is consistent with results indicating that the OS-SQ decomposition yields not only an efficient response basis, but also a forcing basis that is more efficient than the one obtained from the standard resolvent approach \cite{rosenberg2019a}. Furthermore, this estimate was obtained using only information about the velocity statistics, an interesting implication of which is that potentially much can be be learned about the nonlinear forcing directly from commonly-computed flow quantities. We note that similar observations have been made by \citet{towne2020}, who use a limited set of flow statistics to infer forcing statistics, which are in turn used to estimate the unknown flow statistics.

\begin{figure}
	\centering
	\includegraphics[width=0.7\linewidth]{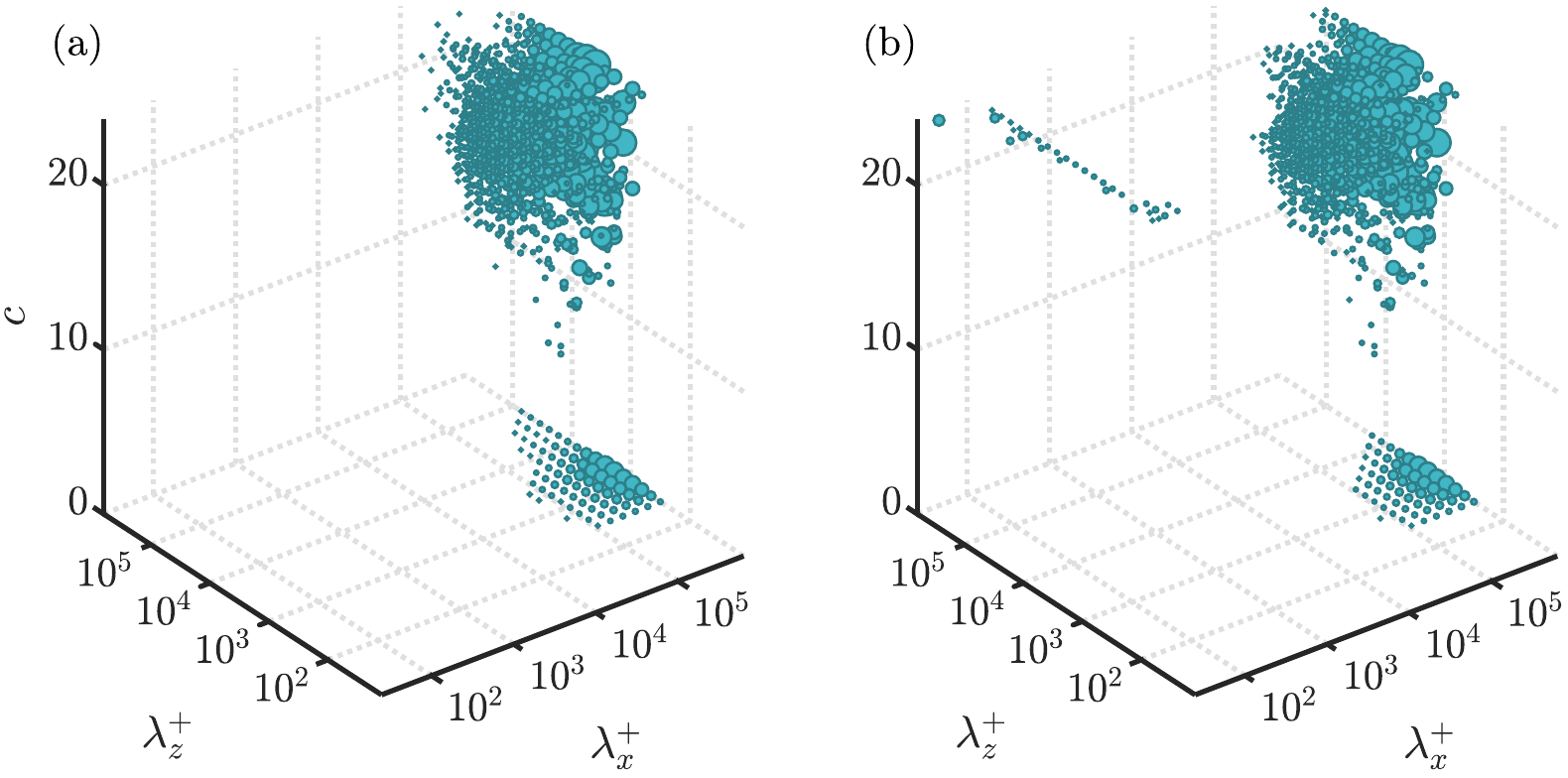}
	\caption{Magnitudes of the total leading mode coefficients (a) $\sigma_1^\OS\sqrt{\tenscomp{X}_{11}^{\OS/\OS}}$, (b) $\sigma_1^\SQ\sqrt{\tenscomp{X}_{11}^{\SQ/\SQ}}$ {larger than 1\% of the maximum value over all of spectral space} for $\Rey=2003$. Marker sizes are proportional to the magnitude and are normalized by the maximum. {The axes show the full range of wave parameters included in the optimization.} \label{fig:pointcloud}}
\end{figure}

Finally, to give additional insight into how energetic modes are distributed across spectral space, the magnitudes of the total mode coefficients, i.e., the weight multiplied by the singular value, are plotted for the leading OS and SQ modes in \cref{fig:pointcloud}(a) and \cref{fig:pointcloud}(b), respectively, for $\Rey=2003$; for ease of visualization, only coefficients larger than 1\% of the maximum value over all spectral space are plotted. Interestingly, they are largely concentrated at large $\lambda_x$ and $c$ close to $U_{cl}$. {In addition, there are large coefficients for very low $c$ and large $\lambda_x^+$, which are likely related to near-wall over-compensation, discussed in \cref{sec:err}.} This observed clustering may have implications for further model reduction by highlighting important regions of spectral space. {Finally, with the exception of some SQ coefficients at small $\lambda_x^+$ and $c\approx U_{cl}$, the large OS and SQ coefficients occupy the essentially the same regions of spectral space which is a reflection of the interactions between the two families of modes; this is discussed in detail in the next section.}

\subsection{Interpretation of the OS-SQ decomposition: a competition mechanism}
It has been demonstrated that the performance of the optimization is greatly improved by employing the OS-SQ decomposition of the resolvent. Previous work reported similar results for channel ECS \cite{rosenberg2018,rosenberg2019a}. In that case, the relatively poor performance of the traditional resolvent method was attributed to the fact that the $\eta$ response dominates under the kinetic energy norm. Thus, matching the statistics for $u$ (or $w$) results in under-prediction of the $v$ statistics, as observed in \cref{fig:kxspect,fig:ints}. However, in the OS-SQ decomposition, isolating the $v$ response in only the OS modes allows $v$ and $\eta$, to be `tuned' somewhat independently, with the role of the SQ modes then being to saturate the OS vorticity. 
The improved matching of all components in \cref{fig:ints} indicates that this is also the case for fully-turbulent channel flow, where the dynamics are significantly more complex than for the aforementioned equilibria. 

\begin{figure}[h]
	\centering                              \includegraphics[width=0.65\textwidth]{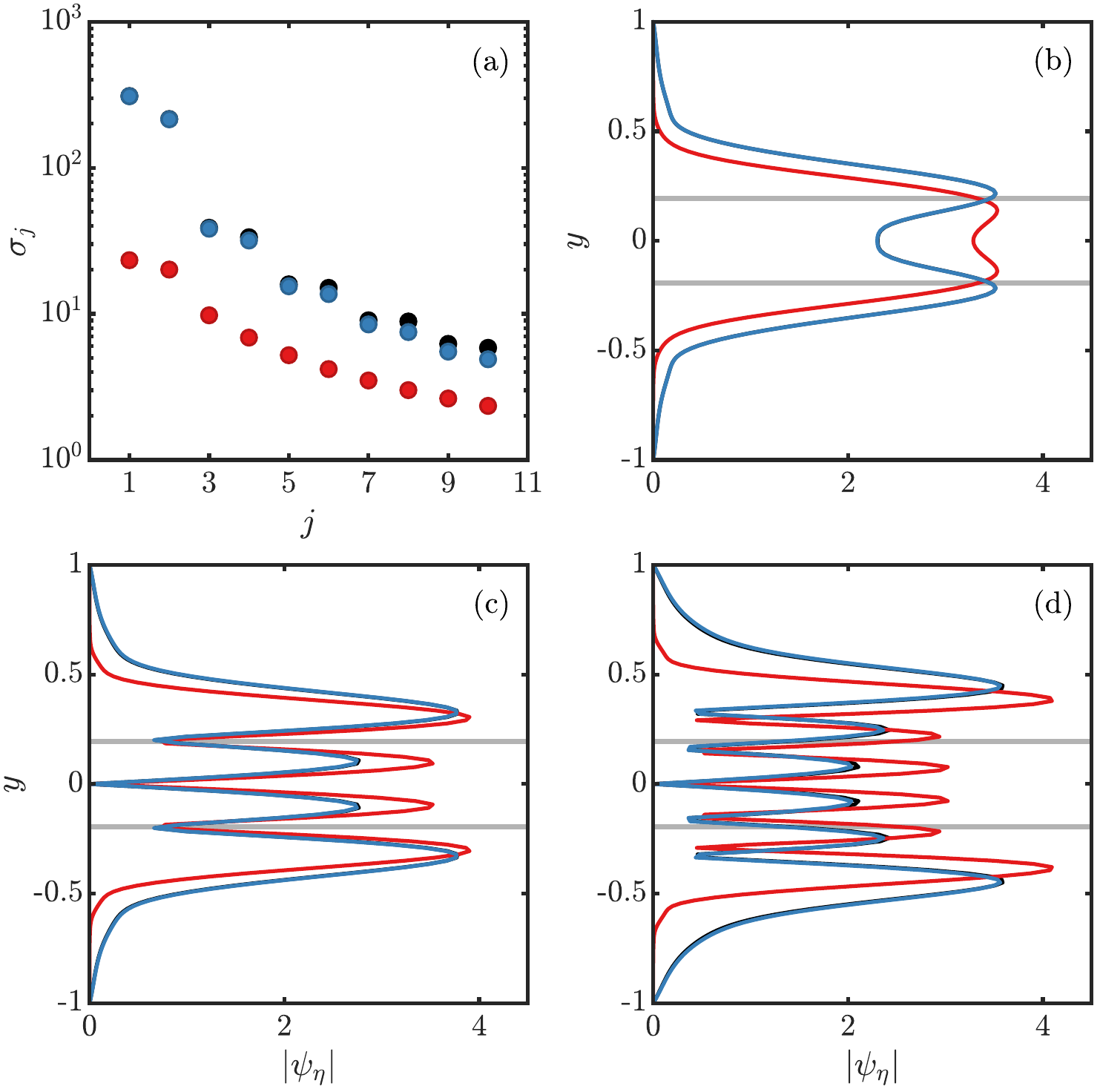}
	\caption{(a) First ten singular values of the OS (blue), SQ (red), and standard (black) resolvent operators for $(k_x,k_z,c)=(0.25, 2.5, 24)$ and $\Rey=2003$. (b)-(d) Magnitudes of the vorticity responses from the first, second, and third mode pairs (same color scheme as in (a)); modes having the same wall-normal symmetry have been selected from each pair. The standard resolvent (black) and OS (blue) modes are visually indistinguishable. The gray lines in (b)-(d) are the locations of the critical layers, $y_c = \pm 0.194$.\label{fig:modes}}
\end{figure}

{To understand why the OS and SQ modes comprise a much more efficient basis, note that in certain cases
the response modes of the standard resolvent, \cref{eq:NSEOS}, coincide with those of the OS resolvent, \cref{eq:resOS}. A detailed description of the regions of parameter space where this holds is beyond the present scope, {but we note that since $\res_{\eta v}$ contains the coupling term $-ik_zU'$, it is expected to hold whenever the lift-up mechanism is dominant, one such example being for highly streamwise-elongated modes.} Further discussion can be found in \citet{dawson2019}.
Here, we simply illustrate by example} for a particular wavenumber-frequency triplet in \cref{fig:modes}, which compares the singular values and magnitudes of the $\hat{\eta}$ response for the standard, OS, and SQ resolvents. Due to the symmetry of the channel geometry about $y=0$, the modes come in symmetric-antisymmetric pairs with correspondingly paired singular values. {The singular values of the OS and standard resolvents are almost equal, with the separation between them growing slowly with increasing mode index.  Looking now at the $\hat{\eta}$ response modes, those from OS and standard resolvents are almost indistinguishable. Though not shown, the same is true for the $\hat{v}$ responses.} 

{The SQ singular values are significantly smaller than those for the standard or OS resolvent -- by more than an order of magnitude for the first pair. Interestingly, the SQ singular values do not demonstrate clear pairing beyond this. The SQ $\hat{\eta}$ modes are distinct from the other two, in particular having slightly narrower wall-normal support. However, the shapes are still largely similar. Importantly, there is still a significant region of overlap with the OS modes in the wall-normal direction, which is a necessary condition for the SQ modes to interact with the the OS modes.} 

\begin{figure}[h]
	\centering                              \includegraphics[width=\textwidth]{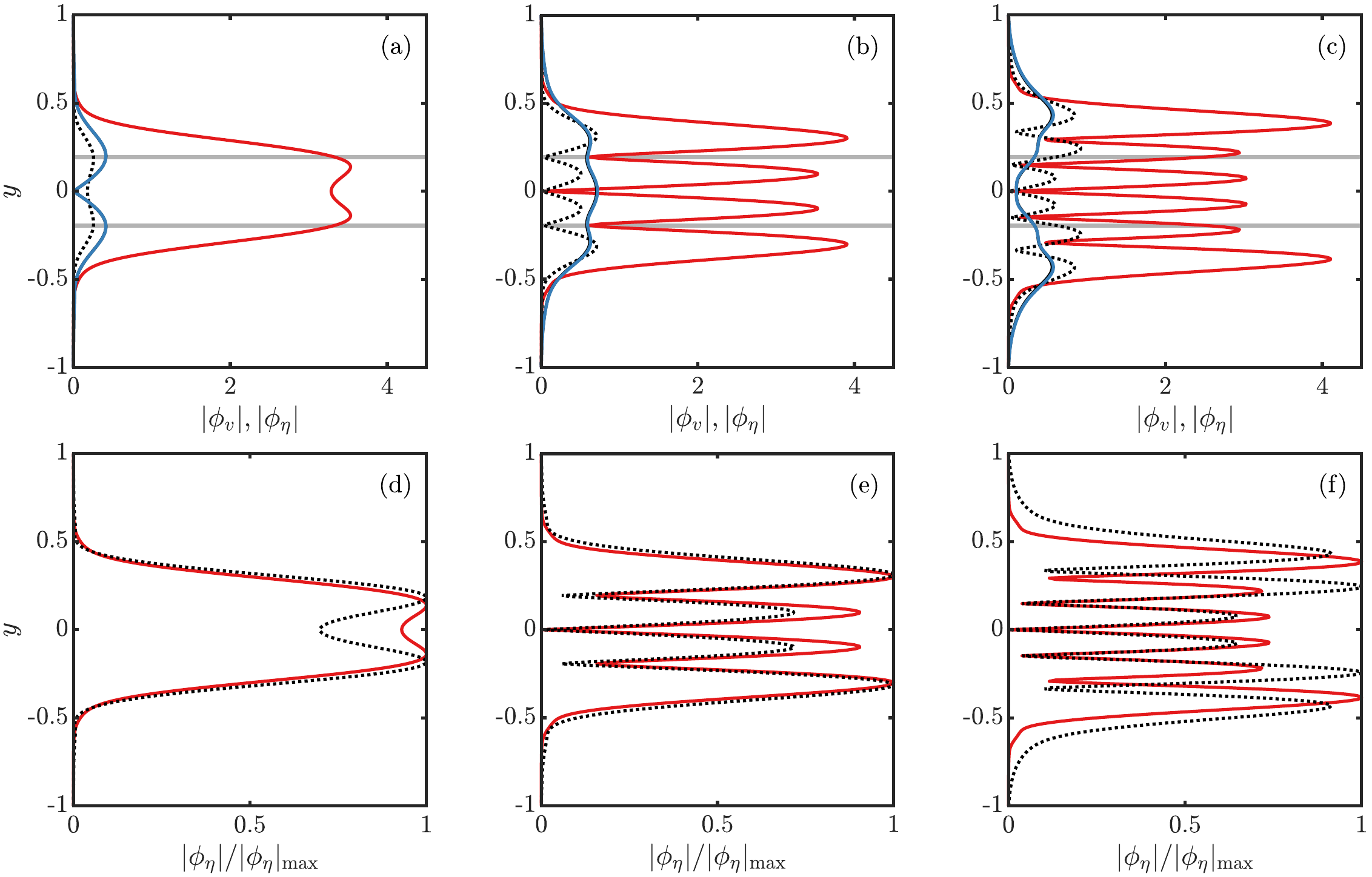}
	\caption{(a)-(c) Magnitudes of the forcing modes corresponding to the response modes shown in \cref{fig:modes}, using the same color scheme. $\phi_v$ for the standard resolvent (solid black) is indistinguishable from OS (blue). Dotted lines are the standard resolvent $\phi_\eta$. (d)-(f) SQ and standard resolvent $\phi_\eta$ normalized by their maximum values. The gray lines in (a)-(c) are the locations of the critical layers, $y_c = \pm 0.194$.\label{fig:forcing}}
\end{figure}

{It is also instructive to look at the corresponding forcing modes, which are shown in the top row of \cref{fig:forcing}. As with the response modes, $\phi_v$ for the OS and standard resolvents are virtually identical. This is at first surprising since the standard resolvent has $\phi_\eta$ with comparable amplitude to $\phi_v$. However, its contribution to the norm is $\lesssim 1\%$. The bottom row of \cref{fig:forcing} shows $\phi_\eta$ for the SQ and standard resolvents normalized by their maximum amplitude for ease of comparison. Despite some differences that become more pronounced for the higher-order modes, their shapes are overall quite similar. Therefore, though there may be traces of the SQ modes in the leading standard resolvent modes, they are clearly dominated by the OS ones. This implies that using the standard resolvent operator to generate a low-order representation is effectively equivalent to using only the OS family of modes, and the linear mechanisms encoded in the SQ operator are thus not accounted for.}

\begin{figure}[h]
	\centering                              \includegraphics[width=\textwidth]{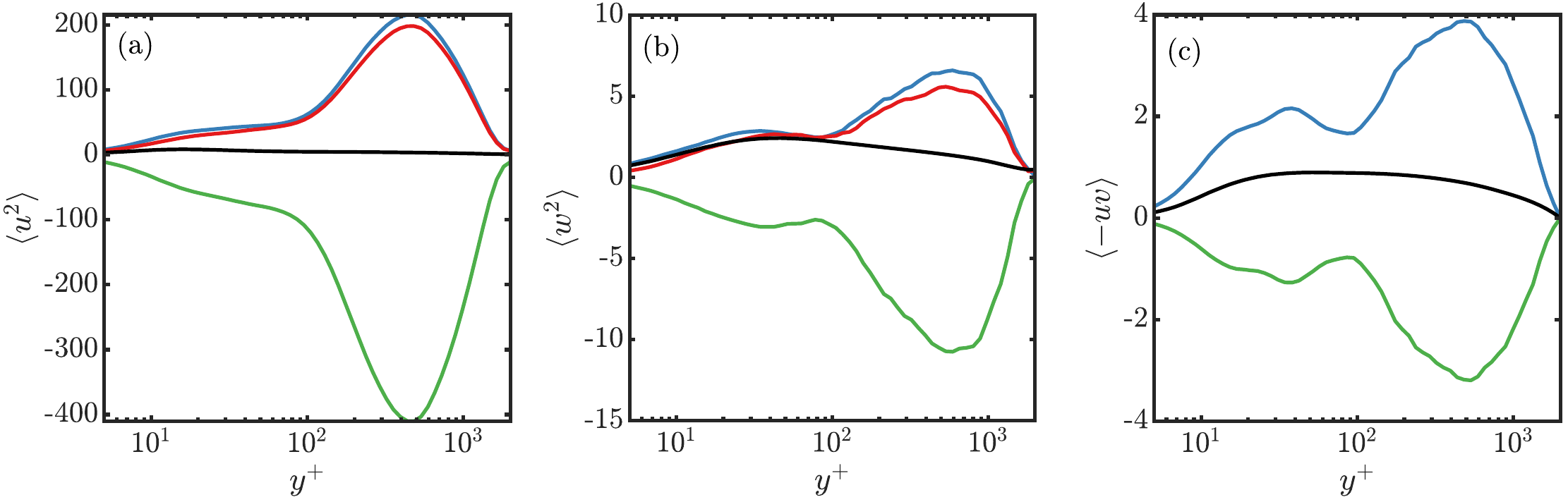}
	\caption{(a) $\langle u^2 \rangle$, (b) $\langle w^2 \rangle$, (c) $\langle -uv \rangle$ decomposed into OS (blue), SQ (red), and C (green) terms for $N=3$. The totals are plotted in black.\label{fig:intsdecomp}}
\end{figure}

To further examine the relationship between the OS and SQ modes, we decompose the intensities shown in \cref{fig:ints} into contributions from OS modes only, SQ modes only, and a cross term (C) that represents the interaction of OS and SQ modes, e.g. $\langle u^2 \rangle$ becomes 
\begin{equation}
{\left\langle u^2 \right\rangle = 
\underbrace{\left\langle \left(u^\OS\right)^2 \right\rangle}_{\OS} 
+ 
\underbrace{\left\langle \left(u^\SQ\right)^2 \right\rangle}_{\SQ} 
+ 
\underbrace{2\left\langle u^\OS u^\SQ \right\rangle}_{\mathrm{C}}.}
\label{eq:udecomp} 
\end{equation}
The results with $N=3$ for $\langle u^2 \rangle$, $\langle w^2 \rangle$, and $\langle -uv \rangle$ are shown in \cref{fig:intsdecomp}. The decomposition for $\langle v^2 \rangle$ is not shown since, as seen from \cref{eq:resSQ}, the SQ modes have no $v$ response, and hence $\langle v^2 \rangle = \langle (v^\OS)^2 \rangle$. Similarly, there is no SQ-only contribution to $\langle -uv \rangle$.
For $\langle u^2 \rangle$ and $\langle w^2 \rangle$, the OS and SQ terms are similar, with the OS term having slightly larger magnitude. However, for all three components the C term is negative, which supports the claim that {the SQ vorticity acts to saturate the OS vorticity.} In fact, information about the phase relationship between the OS and SQ modes can be deduced from this observation. Note that the third term in \cref{eq:udecomp} is simply twice the covariance of $u^\OS$ and $u^\SQ$. {For simplicity, express each as a Fourier sine series in one variable:
\begin{subequations}
\label{eq:fourser}
\begin{align}
u^\OS &= \sum_{i=1}^{\infty} A^\OS_i \sin(k_i x + \theta^\OS_i),
\\
u^\SQ &= \sum_{i=1}^{\infty} A^\SQ_i \sin(k_i x + \theta^\SQ_i),
\end{align}
\end{subequations}
where $A_i^{\mathrm{X}}$ and $\theta_i^{\mathrm{X}}$ are the amplitude and phase of the $i$th mode, respectively, and $0<k_1<k_2<\cdots$. Then,
\begin{align}
2\left\langle u^{\text{OS}}u^{\text{SQ}} \right\rangle &= 2 \sum_i \sum_j A^\OS_i A^\SQ_j \left\langle \sin(k_i x + \theta^\OS_i) \sin(k_j x + \theta^\SQ_j) \right\rangle
\nonumber\\
&=\sum_i \sum_j A^\OS_i A^\SQ_j \left( \left\langle \cos\!\left[ (k_i-k_j) x + \theta^\OS_i-\theta^\SQ_j \right]\right\rangle  - \left\langle \cos\!\left[ (k_i+k_j) x + \theta^\OS_i+\theta^\SQ_j \right]\right\rangle \right)
\label{eq:dtheta}\\
&= \sum_i A^\OS_i A^\SQ_i \cos(\Delta\theta_i),
\nonumber
\end{align}
where $\Delta\theta_i=\theta^\OS_i-\theta^\SQ_i$. The term labeled C in \cref{eq:udecomp} can thus be interpreted as a weighted (by the amplitudes) sum of the cosines of the phase difference between the OS and SQ modes.} Therefore, $\langle u^\OS u^\SQ \rangle < 0$ implies $\pi/2 < \Delta\theta_i < 3\pi/2$ on average. Furthermore, the relative magnitudes of the three terms suggests that for the majority of modes the phase difference is relatively close to $\pi$, i.e, the OS and SQ modes are close to being exactly out of phase. Finally, we note that while the individual terms in \cref{eq:udecomp} depend on $N$, the trends discussed above, namely the similarity of the OS and SQ terms and the C term being negative, do not. Furthermore, performing the decomposition in \cref{eq:udecomp} for $\langle \eta^2 \rangle$ from $\Rey=185$ DNS data reveals the same features \cite{rosenberg2018}. This provides strong evidence that they are not merely consequences of the particular optimization procedure, but are in fact robust features of turbulent channel flow. 

{Since the SQ modes are exclusively wall-parallel motions, there is a passing resemblance to the notion of the `inactive' motions proposed by \citet{townsend1961}. It is supposed that, at first order, the inactive motions to not interact with the `active' shear stress-carrying motions. However, \cref{fig:intsdecomp}(c) shows that the interaction of the SQ vorticity with $v$ produced by the OS modes, the C term, contributes significantly to the overall Reynolds shear stress profile, suggesting that there is not an exact correspondence between the SQ modes and inactive motions.}\

{In this section it was shown that the OS-SQ decomposition of the resolvent provides an improved basis for efficiently representing the statistics of turbulent channel flow, and that this provides insight into the complex physics at play, namely a competition mechanism, interpreted as a phase difference, between OS and SQ modes that results in saturation of the wall-normal vorticity. In the next section, we use this insight to derive simple scalings for the relative magnitudes of the OS and SQ weights of modes belonging to several special classes.}

\section{\label{sec:scaling}Weights scaling for the universal classes of resolvent modes}
\citet{moarref2013} leveraged universal scaling regimes of the mean velocity profile to derive the $\Rey$ scaling for several universal classes of resolvent modes. Here, we extend this to the OS-SQ resolvent decomposition and show that for the outer and geometrically self-similar classes, each family of modes has a distinct scaling for the singular values. 
From this, the scalings of each submatrix of the energy density matrices $\tens{A}_r$ given in \cref{eq:partition} can be determined. Combining these scalings with the hypothesis that competition of the OS and SQ modes remains relevant at different Reynolds numbers and in different regions of the flow enables the relative scaling of the OS and SQ weights belonging to the universal classes to be deduced. 

\begin{figure}[h]
	\centering \includegraphics[width=0.7\textwidth]{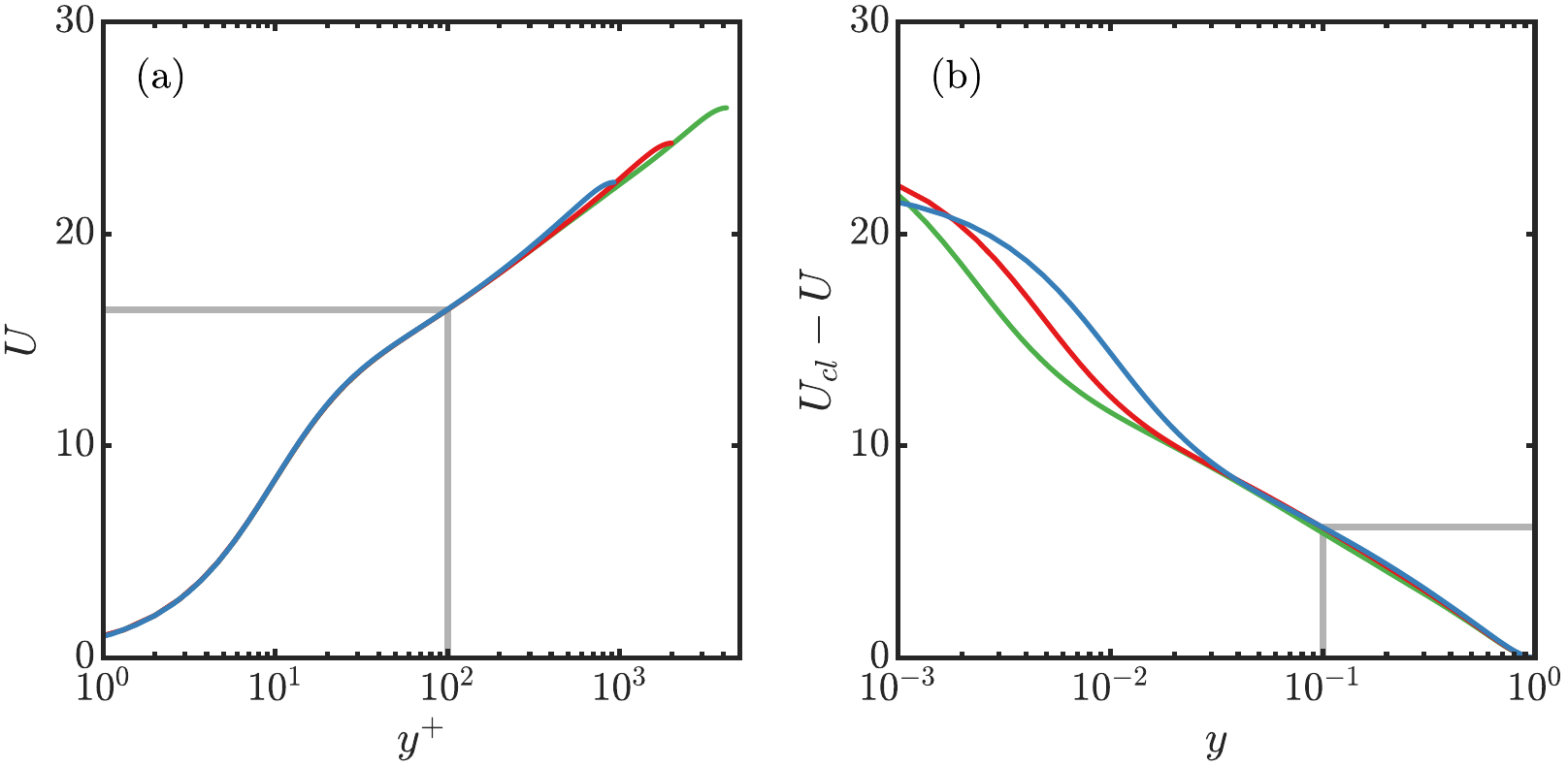}
	\caption{(a) Mean velocity profile $U(y^+)$ and (b) velocity defect $U_{cl}-U(y)$ for $\Rey=934$ (blue), $\Rey=2003$ (red), and $\Rey=4219$ (green). The gray boxes indicate the regions where the profiles are $\Rey$-invariant in the respective coordinates. \label{fig:means}}
\end{figure}

The universal classes investigated here are the inner, outer, and geometrically self-similar classes. These consist of resolvent modes that are localized within the near-wall, wake, and logarithmic regions of the flow, respectively, and rely on universality of the mean velocity profile under the appropriate scaling in these regions. \cref{fig:means}(a) demonstrates the universality of $U$ for $y^+\lesssim100$, and \cref{fig:means}(b) shows that the velocity defect $U_{cl}-U$ is universal for $y\gtrsim0.1$; these approximate boundaries are indicated by the vertical dashed lines in \cref{fig:means}. {Between these regions, there exists an intermediate region of the mean velocity profile in which both scalings hold. In this overlap region, it is widely accepted that the mean varies logarithmically with distance from the wall.} Classical estimates put the beginning of the logarithmic region at $y^+=O(100)$. However, there is recent evidence that this lower boundary moves outward as $\Rey^{1/2}$ \cite{klewicki2009,marusic2013}. In the logarithmic region, the resolvent operator admits self-similar modes localized about their critical layers \cite{moarref2013}. {Furthermore, the scaling of these modes reduces to the inner and outer scalings when $y^+$ or $y$, respectively, is held fixed, reflecting their mutual validity in the logarithmic region.}

In each of the next subsections, we briefly summarize the scaling of the wave parameters for each of the three aforementioned universal classes derived by \citet{moarref2013}, as well as the distinct scalings for the OS and SQ singular values. Finally, the relative scaling of the OS and SQ weights are presented and tested against the computed optimal weights.

\subsection{\label{sec:inner}Inner class}
\begin{figure}[h]
	\centering \includegraphics[width=0.7\textwidth]{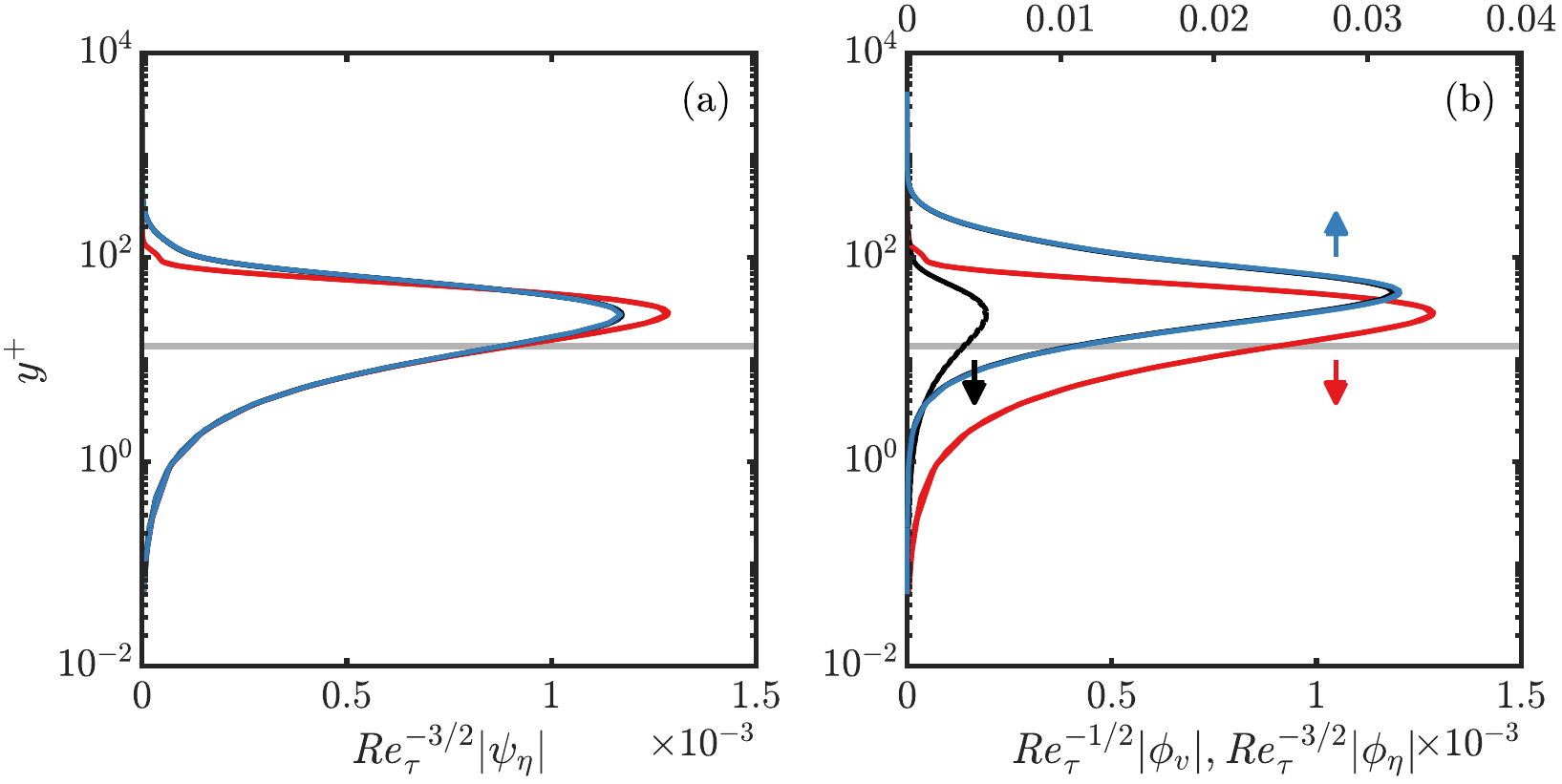}
	\caption{(a) Magnitudes of the scaled OS, SQ, and standard resolvent (color scheme as in~\cref{fig:modes}) inner-class leading vorticity response modes for the three $\Rey$ shown in~\cref{fig:means}, with $k_x^+ = 1/934$, $k_z^+ = 10 k_x^+$, and $c=10$. (b) Corresponding scaled leading forcing mode magnitudes, with the relevant axes indicated by the arrows. $\psi_\eta$ and $\phi_v$ for the standard resolvent are indistinguishable from the OS modes. The gray line indicates the location of the critical layer. \label{fig:modesinner}}
\end{figure}

Following \citet{moarref2013}, the relevant length scale for the inner class is the viscous unit $\nu/u_\tau$, so that the corresponding inner-scaled parameters are
\begin{equation}
k_x^+ = \Rey^{-1}k_x, \quad k_z^+ = \Rey^{-1}k_z, \quad y^+ = \Rey y,
\label{eq:li}
\end{equation}
and the inner class wave parameters are
\begin{equation}
\mathscr{S}_i: \quad 0 \leq c \lesssim 16.4.
\label{eq:Si}
\end{equation}
The upper wavespeed limit is obtained from the critical layer at the top of the inner region, i.e., $U(y^+=100)=16.4$; this is indicated by the horizontal dashed line in \cref{fig:means}a. Note that this bound is slightly different from the one given in \citet{moarref2013} since the mean velocity profiles they used were obtained from an eddy viscosity model, whereas the ones used here are taken directly from the DNS that the spectra are obtained from. Using \cref{eq:li} and continuity, it follows that all three velocity components scale in the same way. Furthermore, the orthonormality constraint on the resolvent modes imposes 
\begin{equation}
\hat{\vect{u}} = \Rey^{1/2}\hat{\vect{u}}^+,
\label{eq:ui}
\end{equation}
where a superscript $\blankop^+$ indicates a quantity that is $\Rey$-invariant for modes belonging to the inner class. 
\cref{eq:li} can be used to obtain the inner-scaled versions of the weighted resolvent operators:
\begin{subequations}
	\label{eq:Hopsi}
	\begin{align}
	\begin{pmatrix}
	\op{F}_v \res_{vv} \op{F}_v^\inv \\
	\op{F}_\eta \res_{\eta v} \op{F}_v^\inv
	\end{pmatrix} 
	&= \Rey^\inv 
	\begin{pmatrix}
	\op{F}_v^+ \res_{vv}^+ \op{F}_v^{+\, \inv} \\
	\op{F}_\eta^+ \res_{\eta v}^+ \op{F}_v^{+\, \inv}
	\end{pmatrix}
	\label{eq:Hosi}
	\\
	\op{F}_\eta \res_{\eta\eta} \op{F}_\eta^\inv &= \Rey^\inv \op{F}_\eta^+ \res_{\eta\eta}^+ \op{F}_\eta^{+\, \inv},
	\label{eq:Hsqi}
	\end{align}
\end{subequations}
where $\op{F} = \diag( \op{F}_v, \, \op{F}_\eta)$ is the square root of the positive-definite operator $\op{Q}$ defined in \cref{eq:Q}, i.e., $\op{Q} = \op{F}^{\adj}\op{F}$, and the superscript $\blankop^\adj$ denotes the adjoint with respect to the inner product \cref{eq:IP}. Computing the SVDs of \cref{eq:Hosi,eq:Hsqi}, it is clear that both the OS and SQ singular values have the same scaling:
\begin{equation}
\sigma_j^\OS = \Rey^\inv \sigma_j^{\OS\, +}, \quad \sigma_j^\SQ = \Rey^\inv \sigma_j^{\SQ\, +}.
\label{eq:sigi}
\end{equation}
{For a particular $k_x^+$, $k_z^+$, and $c\in\mathscr{S}_i$, the magnitudes of the leading vorticity response modes of the OS and SQ resolvent operators and their corresponding forcing modes for the three $\Rey$ in \cref{fig:means} are shown in \cref{fig:modesinner}(a) and \cref{fig:modesinner}(b), respectively, using the scalings derived in \crefrange{eq:ui}{eq:sigi}. For comparison, the leading response and forcing mode for the standard resolvent operator are also shown; as discussed in \cref{sec:kxspectra}, $\psi_\eta$ and $\phi_v$ are indistinguishable from the OS modes.}

Substituting \cref{eq:ui,eq:sigi} into \cref{eq:A}, we obtain the inner-scaled energy density matrices:
\begin{equation}
	\begin{array}{c c c}
	\tenscomp{A}^{\OS/\OS}_{r,ij} = \Rey^\inv \tenscomp{A}^{\OS/\OS\, +}_{r,ij}, &
	\tenscomp{A}^{\OS/\SQ}_{r,ij} = \Rey^\inv \tenscomp{A}^{\OS/\SQ\, +}_{r,ij}, &
	\tenscomp{A}^{\SQ/\SQ}_{r,ij} = \Rey^\inv \tenscomp{A}^{\SQ/\SQ\, +}_{r,ij}.
	\end{array}
	\label{eq:Ai}
\end{equation}
With this, the decomposed version of the three-dimesional streamwise energy spectrum becomes 
\begin{equation}
E_{uu} = \Rey^\inv \Re\!\left\{ \tr\! \left( \tens{A}^{\OS/\OS\, +}_{uu}\tens{X}^{\OS/\OS} \right) \right\} + 2\Rey^\inv \Re\!\left\{ \tr\! \left( \tens{A}^{\SQ/\OS\, +}_{uu}\tens{X}^{\OS/\SQ} \right) \right\} + \Rey^\inv \Re\!\left\{ \tr\! \left( \tens{A}^{\SQ/\SQ\, +}_{uu}\tens{X}^{\SQ/\SQ} \right) \right\},
\label{eq:Euui} 
\end{equation} 
where the Reynolds number dependence  of the right-hand side is made explicit, save for the unscaled weights matrices $\tens{X}^{\text{X}/\text{Y}}$. \cref{eq:Ai} can be used to write similar expressions for the other components of the spectra. 

Since the overall scaling of $E_{uu}$ for the inner class is not known, the absolute scaling of the weights cannot be determined directly from \cref{eq:Euui}. However, in \cref{sec:kxspectra} it was shown that the vorticity generated by OS and SQ modes compete. That is, the vorticity generated by the SQ modes acts to `saturate' the OS vorticity. We hypothesize that this mechanism is not specific to $\Rey=2003$ for which the optimization results were presented, but instead holds for arbitrary $\Rey$. This is only possible if all three terms remain of the same order in \cref{eq:Euui}, which is satisfied if the inner class OS and SQ weights have the same scaling, i.e., if the ratio 
\begin{equation}
\left\vert \frac{\chi_j^\SQ}{\chi_j^\OS} \right\vert \neq \mathrm{fn}(\Rey)
\label{eq:rati}
\end{equation} 
for modes belonging to the universal inner class.

\begin{figure}[h]
	\centering \includegraphics[width=\textwidth]{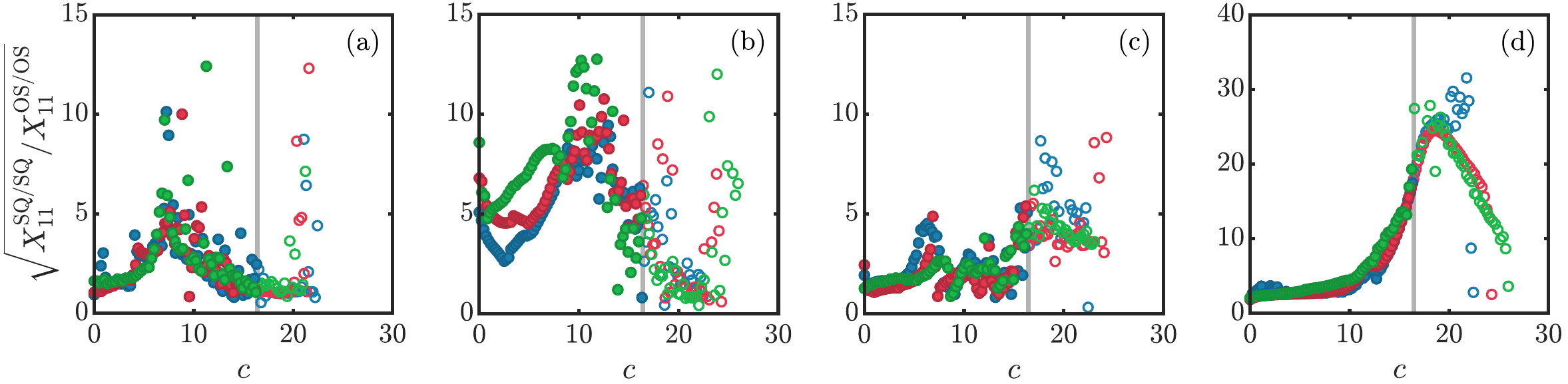}
	\caption{Leading weights ratio for $\Rey=934$ (blue), $\Rey=2003$ (red), and $\Rey=4219$ (green) with $N=3$ for several different wavelengths: (a) $(k_x^+, k_z^+)=(2\pi/10^2,2\pi/10^2)$; (b) $(k_x^+, k_z^+)=(2\pi/10^3,2\pi/10^2)$; (c) $(k_x^+, k_z^+)=(2\pi/10^3,2\pi/10^3)$; (d) $(k_x^+, k_z^+)=(2\pi/10^4,2\pi/10^3)$. Filled circles denote modes belonging to the universal inner class, and the highest inner class wavespeed, $c = 16.4$, is indicated by the gray line.\label{fig:win}}
\end{figure}

This scaling is tested by computing the weights matrices for the three Reynolds numbers depicted in \cref{fig:means} for fixed inner-scaled wavenumber combinations $(k_x^+,k_z^+)$. As discussed in \cref{sec:opt}, the individual weights are not recovered from the full-rank solutions. However, 
$\tenscomp{X}_{jj}^{\OS/\OS} \sim \vert \chi^\OS_j \vert^2$ and $\tenscomp{X}_{jj}^{\SQ/\SQ} \sim \vert \chi^\SQ_j \vert^2$, so that $\vert\chi_j^\SQ/\chi_j^\OS\vert \sim \sqrt{\tenscomp{X}_{jj}^{\SQ/\SQ}/\tenscomp{X}_{jj}^{\OS/\OS}}$. This ratio with $j=1$ is computed for the three Reynolds numbers, and the results for several wavenumber combinations spanning a large range of scales are shown in \cref{fig:win}. Overall, the agreement is quite good, showing a reasonable collapse despite some scatter. The main exception is for $(k_x^+, k_z^+)=(2\pi/10^3,2\pi/10^2)$ in \cref{fig:win}b, where there is clearly some dependence on $\Rey$ for $c\lesssim10$. This trend is consistent for modes of relatively small scale ($k_x^+ \gtrsim O(2\pi/10^3)$) and large aspect ratio ($k_z^+/k_x^+ \gtrsim O(10)$). The reason for the failure of the scaling for these modes is unclear. However, it is observed that in such cases the profiles of the time-averaged energy spectra are localized very near the wall. Additionally, the upper limit on the wavespeed range for inner class modes given in \cref{eq:Si}, $c=16.4$ is shown in each panel of \cref{fig:win} as the vertical gray line. {The fact that the scaling given by \cref{eq:rati} holds reasonably well for $c > 16.4$ is a reflection of the fact that the inner scaling remains valid in the logarithmic region, as seen in \cref{fig:means}(a).}

\subsection{\label{sec:outer}Outer class}
The outer class length scales are
\begin{equation}
\out{k}_x = \Rey k_x, \quad \out{k}_z = k_z, \quad \out{y} = y,
\label{eq:lo}
\end{equation}
and the outer class wave parameters are
\begin{equation}
\mathscr{S}_{o} : \left\{ 
\begin{array}{l}
	0 \leq U_{c l}-c \lesssim 6.17 
	\\ 
	k_z / k_x \gtrsim \gamma \Rey / \mathit{Re}_{\tau, \min}
	\end{array} \right.,
\label{eq:So}
\end{equation}
where The upper bound on the wavespeed defect $U_{cl}-c=6.17$ is obtained from the setting the minimum critical layer location at the bottom of the outer region, i.e., $U_{cl}-U(y=0.1)=6.17$; this is indicated by the horizontal dashed line in \cref{fig:means}b. Again, this value is slightly different from the one given in \citet{moarref2013} due to the different source for the mean profiles. As \cref{eq:So} indicates, the outer class modes must satisfy an aspect ratio constraint for all $\Rey$ considered, where the minimum aspect ratio is $\gamma$ when $\Rey=\mathit{Re}_{\tau,\min}$ \cite{moarref2013}. Here, $\mathit{Re}_{\tau,\min}=934$. From \cref{eq:lo} and  continuity, it follows that 
\begin{equation}
\hat{\vect{u}} =
\begin{pmatrix}
\out{u}\\
\Rey^\inv\out{v}\\
\Rey^\inv\out{w}
\end{pmatrix}. 
\label{eq:uo}
\end{equation}
where $\out{\blankop}$ indicates a quantity that is approximately $\Rey$-invariant for modes belonging to the outer class. {See also \citet{sharma2017} and \citet{moarref2014b} for the scaling of each velocity component, as well as for the components of the forcing modes.}
\begin{figure}[h]
	\centering \includegraphics[width=0.7\textwidth]{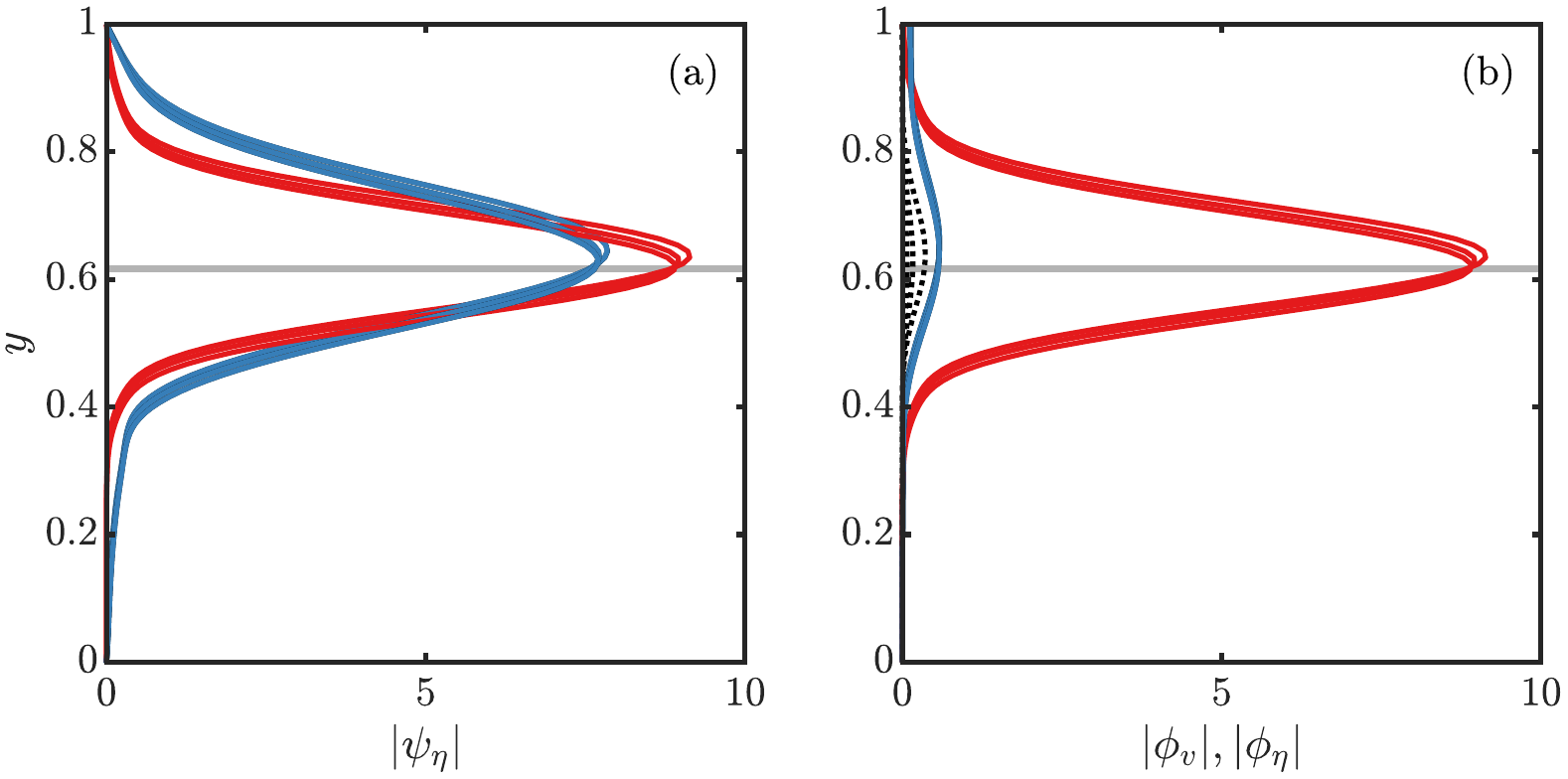}
	\caption{(a) Magnitudes of the scaled OS, SQ, and standard resolvent (color scheme as in~\cref{fig:modes}) outer-class leading vorticity response modes for the three $\Rey$ shown in~\cref{fig:means}, with $\out{k}_x = 934$, $\out{k}_z = \gamma=1.5\sqrt{10}$, and $U_{cl}-c=1$. (b) Corresponding scaled leading forcing mode magnitudes. $\psi_\eta$ and $\phi_v$ for the standard resolvent are indistinguishable from the OS modes. The gray line indicates the location of the critical layer. \label{fig:modesouter}}
\end{figure}  

The outer-scaled versions of the weighted resolvent operators are
\begin{subequations}
	\label{eq:Hopso}
	\begin{align}
	\begin{pmatrix}
	\op{F}_v \res_{vv} \op{F}_v^\inv \\
	\op{F}_\eta \res_{\eta v} \op{F}_v^\inv
	\end{pmatrix} 
	&=  
	\begin{pmatrix}
	\Rey \out{\op{F}}_v \out{\res}_{vv} \out{\op{F}}_v^\inv \\
	\Rey^2 \out{\op{F}}_\eta \out{\res}_{\eta v} \out{\op{F}}_v^\inv
	\end{pmatrix}
	\label{eq:Hoso}
	\\
	\op{F}_\eta \res_{\eta\eta} \op{F}_\eta^\inv &= \Rey \out{\op{F}}_\eta \out{\res}_{\eta\eta} \out{\op{F}}_\eta^\inv.
	\label{eq:Hsqo}
	\end{align}
\end{subequations}
Computing the SVDs of \cref{eq:Hoso,eq:Hsqo}, we have for the leading singular values,
\begin{equation}
\sigma_j^\OS = \Rey^2 \out{\sigma}_j^\OS, \quad \sigma_j^\SQ = \Rey \out{\sigma}_j^\SQ.
\label{eq:sigo}
\end{equation}
Note that because the components of \cref{eq:Hoso} do not scale uniformly for outer class modes, the scaling of the OS singular values is only expected to hold for the first several modes. However, since good agreement between the resolvent and DNS spectra is achieved using only a small number of modes, it is reasonable to adopt the scalings in what follows. 

{For a particular $(k_x, k_z, c)\in\mathscr{S}_o$, the magnitudes of the leading vorticity response modes of the OS and SQ resolvent operators and their corresponding forcing modes for the three $\Rey$ in \cref{fig:means} are shown in \cref{fig:modesouter}(a) and \cref{fig:modesouter}(b), respectively, using the scalings derived in \crefrange{eq:uo}{eq:sigo}. For comparison, the leading response and forcing mode for the standard resolvent operator are also shown; again, $\psi_\eta$ and $\phi_v$ are indistinguishable from the OS modes. Apparent from \cref{fig:modesouter} is that the scaling of the outer class modes is only approximate. Recalling that the derivation of such universal classes relies on universal behavior of the mean profile, this is not surprising, since it is clear from \cref{fig:means}(b) that the mean profiles for the three $\Rey$ do not collapse perfectly for $y>0.1$. Additionally, $\phi_\eta$ for the standard resolvent (black dotted line in \cref{fig:modesouter}(b)) does not obey the same scaling as $\phi_\eta^\SQ$. Indeed, using the scaling of the standard resolvent in primitive variables presented in \citet{sharma2017}, it can be shown that $\phi_\eta = O(\Rey^\inv)$.}

Substituting \cref{eq:uo,eq:sigo} into \cref{eq:A}, we obtain the outer-scaled energy density matrices:
\begin{subequations}
	\begin{align}
	&\begin{array}{c c c}
	\tenscomp{A}^{\OS/\OS}_{uu,ij} = \Rey^4 \out{\tenscomp{A}}^{\OS/\OS}_{uu,ij}, &
	\tenscomp{A}^{\OS/\SQ}_{uu,ij} = \Rey^3 \out{\tenscomp{A}}^{\OS/\SQ}_{uu,ij}, &
	\tenscomp{A}^{\SQ/\SQ}_{uu,ij} = \Rey^2 \out{\tenscomp{A}}^{\SQ/\SQ}_{uu,ij},
	\end{array}
	\\
	&\begin{array}{c}
	\tenscomp{A}^{\OS/\OS}_{vv,ij} = \Rey^2 \out{\tenscomp{A}}^{\OS/\OS}_{vv,ij},
	\end{array}
	\\
	&\begin{array}{c c c}
	\tenscomp{A}^{\OS/\OS}_{ww,ij} = \Rey^2 \out{\tenscomp{A}}^{\OS/\OS}_{ww,ij}, &
	\tenscomp{A}^{\OS/\SQ}_{ww,ij} = \Rey \out{\tenscomp{A}}^{\OS/\SQ}_{ww,ij}, &
	\tenscomp{A}^{\SQ/\SQ}_{ww,ij} =  \out{\tenscomp{A}}^{\SQ/\SQ}_{ww,ij},
	\end{array}
	\\
	&\begin{array}{c c}
	\tenscomp{A}^{\OS/\OS}_{uv,ij} = \Rey^3 \out{\tenscomp{A}}^{\OS/\OS}_{uv,ij}, &
	\tenscomp{A}^{\OS/\SQ}_{uv,ij} = \Rey^2 \out{\tenscomp{A}}^{\OS/\SQ}_{uv,ij}.
	\end{array}
	\end{align}
	\label{eq:Ao}
\end{subequations}
The streamwise energy spectrum is thus
\begin{equation}
E_{uu} = \Rey^4 \,\Re\!\left\{ \tr\! \left( \out{\tens{A}}^{\OS/\OS}_{uu}\tens{X}^{\OS/\OS} \right) \right\} + 2\Rey^3 \,\Re\!\left\{ \tr\! \left( \out{\tens{A}}^{\SQ/\OS}_{uu}\tens{X}^{\OS/\SQ} \right) \right\} + \Rey^2 \,\Re\!\left\{ \tr\! \left( \out{\tens{A}}^{\SQ/\SQ}_{uu}\tens{X}^{\SQ/\SQ} \right) \right\},
\label{eq:Euuo} 
\end{equation}
where again the Reynolds number dependence of the right-hand side is explicit, save for the unscaled weights matrices $\tens{X}^{\text{X}/\text{Y}}$, and \cref{eq:Ao} can be used to write similar expressions for the other energy spectra.
 
As for the inner class modes, competition between the OS and SQ modes requires that all three terms are of the same order for arbitrary $\Rey$, which is satisfied if
\begin{equation}
\left\vert \frac{\chi_j^\SQ}{\chi_j^\OS} \right\vert \sim \Rey
\label{eq:rato}
\end{equation} 
for modes belonging to the universal outer class.

\begin{figure}[h]
	\centering \includegraphics[width=\textwidth]{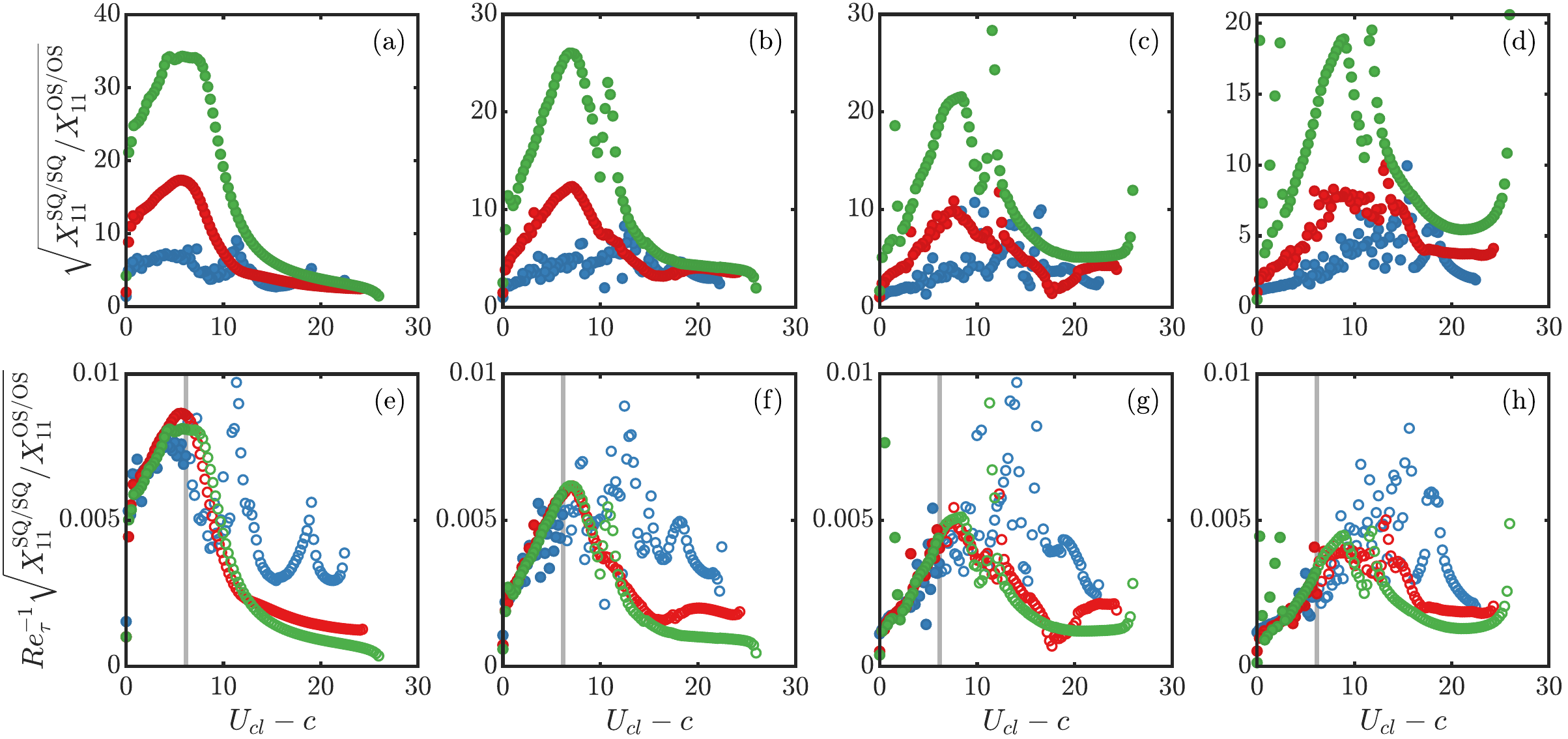}
	\caption{(a)-(d) Leading weights ratio for $\Rey=934$ (blue), $\Rey=2003$ (red), and $\Rey=4219$ (green) with $N=2$. (e)-(h) Weights ratio scaled according to \cref{eq:rato}. Each column represents a different scaled streamwise wavenumber $\out{k}_x$: (a),(e) $\out{k}_x=4219$; (b),(f) $\out{k}_x=8438$; (c),(g) $\out{k}_x=12657$; (d),(h) $\out{k}_x=16876$. In all cases the spanwise wavenumber is $k_z=\gamma \out{k}_x/\mathit{Re}_{\tau, \min}$, with $\gamma=\sqrt{10}$. Filled circles denote modes belonging to the universal outer class, and the largest outer class wavespeed defect, $U_{cl}-c = 6.17$, is indicated by the gray line.\label{fig:wout}}
\end{figure}

The weights ratio $\sqrt{\tenscomp{X}_{11}^{\SQ/\SQ}/\tenscomp{X}_{11}^{\OS/\OS}}$ for the three Reynolds numbers is shown in the top row of \cref{fig:wout} for several values of the outer-scaled wavenumber combinations $(\out{k}_x,\out{k}_x)$ and a minimum aspect ratio $\gamma=\sqrt{10}$. In agreement with \cref{eq:rato}, the data from all three Reynolds numbers show reasonable collapse onto a single curve for $U_{cl}-c\lesssim 6.17$ when scaled by $\Rey^\inv$, as seen in the bottom row of \cref{fig:wout}.

\subsection{\label{sec:selfssim}Geometrically self-similar class}
\begin{figure}[h]
	\centering \includegraphics[width=\textwidth]{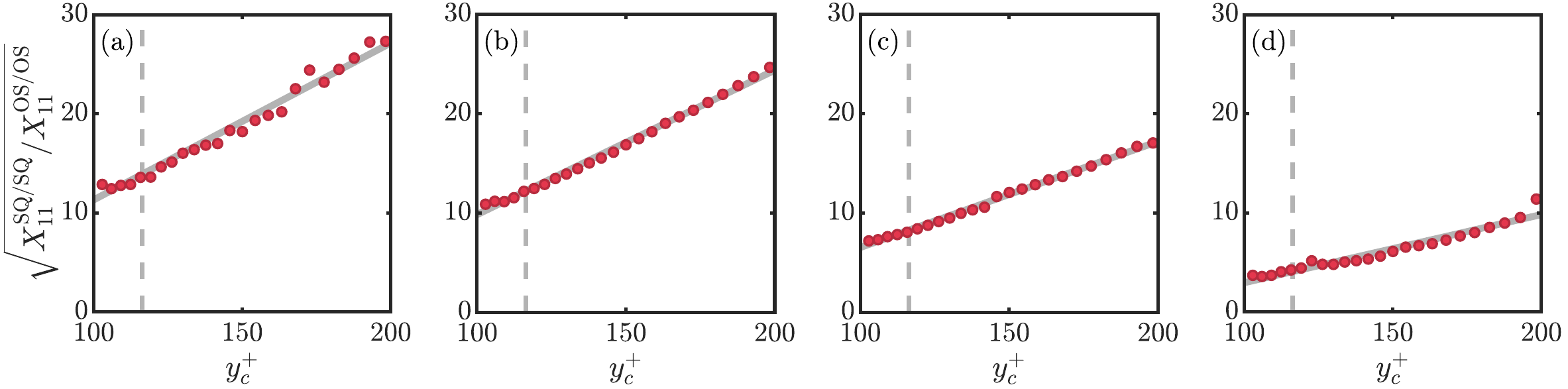}
	\caption{Leading weights ratio along hierarchies for $\Rey=2003$ with $N=2$. Each panel is a different hierarchy, represented by the streamwise wavenumber at the bottom of the hierarchy: (a) $k_{x,l}=1$; (b) $k_{x,l}=5$; (c) $k_{x,l}=10$; (d) $k_{x,l}=20$. In all cases the spanwise wavenumber at the bottom of the hierarchy is $k_{z,l}=\gamma k_{x,l}$, with $\gamma=5$. {The solid gray lines are the least squares linear fits, with slopes (a) 0.158, (b) 0.145, (c) 0.107, and (d) 0.069.} The dashed gray lines are $y^+=2.6\Rey^{1/2}\approx116$.\label{fig:wss}}
\end{figure}

The self-similar resolvent modes in the logarithmic region of the mean velocity profile belong to hierarchies parameterized by the critical layer location $\yc$ \cite{moarref2013}. The corresponding length scales along a hierarchy are
\begin{equation}
\ssim{k}_x = \ycp\yc k_x, \quad \ssim{k}_z = \yc k_z, \quad \ssim{y} = y/\yc,
\label{eq:lss}
\end{equation}
and the self-similar class wave parameters are
\begin{equation}
\mathscr{S}_{h} : \left\{ 
\begin{array}{l}
16.4 \lesssim c \lesssim U_{c l}-6.17 
\\
c = U(\ycp) = \kappa^\inv \log \ycp + B
\\ 
k_z / k_x \gtrsim \gamma 
\end{array} \right.,
\label{eq:Sss}
\end{equation}
where $\kappa$ is the K\'{a}rm\'{a}n constant. Note that the lower wavespeed bound for the self-similar class, $c=16.4$, is the same as the upper limit for the inner class modes, i.e., the beginning of the logarithmic region is taken to be $y^+=100$. As discussed above, recent evidence suggests that this lower limit is $\Rey$-dependent \cite{klewicki2009,marusic2013}. However, \citet{moarref2013} demonstrated successful scaling of the self-similar modes using the fixed lower limit in \cref{eq:Sss}, so we continue to use it here. For reference, the beginning of the logarithmic region according to the balance of terms in the mean momentum equation, $y^+\approx2.6\Rey^{1/2}$ \cite{klewicki2009}, is indicated by the gray dashed line in \cref{fig:wss}. The self-similar modes must also satisfy an aspect ratio constraint. Since the aspect ratio increases like $\ycp$ along a given hierarchy, it is sufficient that the lowest member on the hierarchy with critical layer $y_{c,l}$ and wavenumbers $k_{x,l},k_{z,l}$ satifies $k_{z,l} / k_{x,l} \gtrsim \gamma$, where a conservative lower bound is $\gamma\approx\sqrt{10}$ \cite{moarref2013}.

Using \cref{eq:lss} and continuity, as well as the orthonormality constraint on the resolvent modes, it follows that 
\begin{equation}
\hat{\vect{u}} = \yc^{-1/2}
\begin{pmatrix}
\ssim{u}\\
\yc^{+\, \inv} \ssim{v}\\
\yc^{+\, \inv} \ssim{w}
\end{pmatrix}. 
\label{eq:uss}
\end{equation}
where $\ssim{\blankop}$ indicates a quantity that is approximately $\yc$- and $\Rey$-invariant for modes belonging to the self-similar class. 
\begin{figure}[h]
	\centering \includegraphics[width=0.7\textwidth]{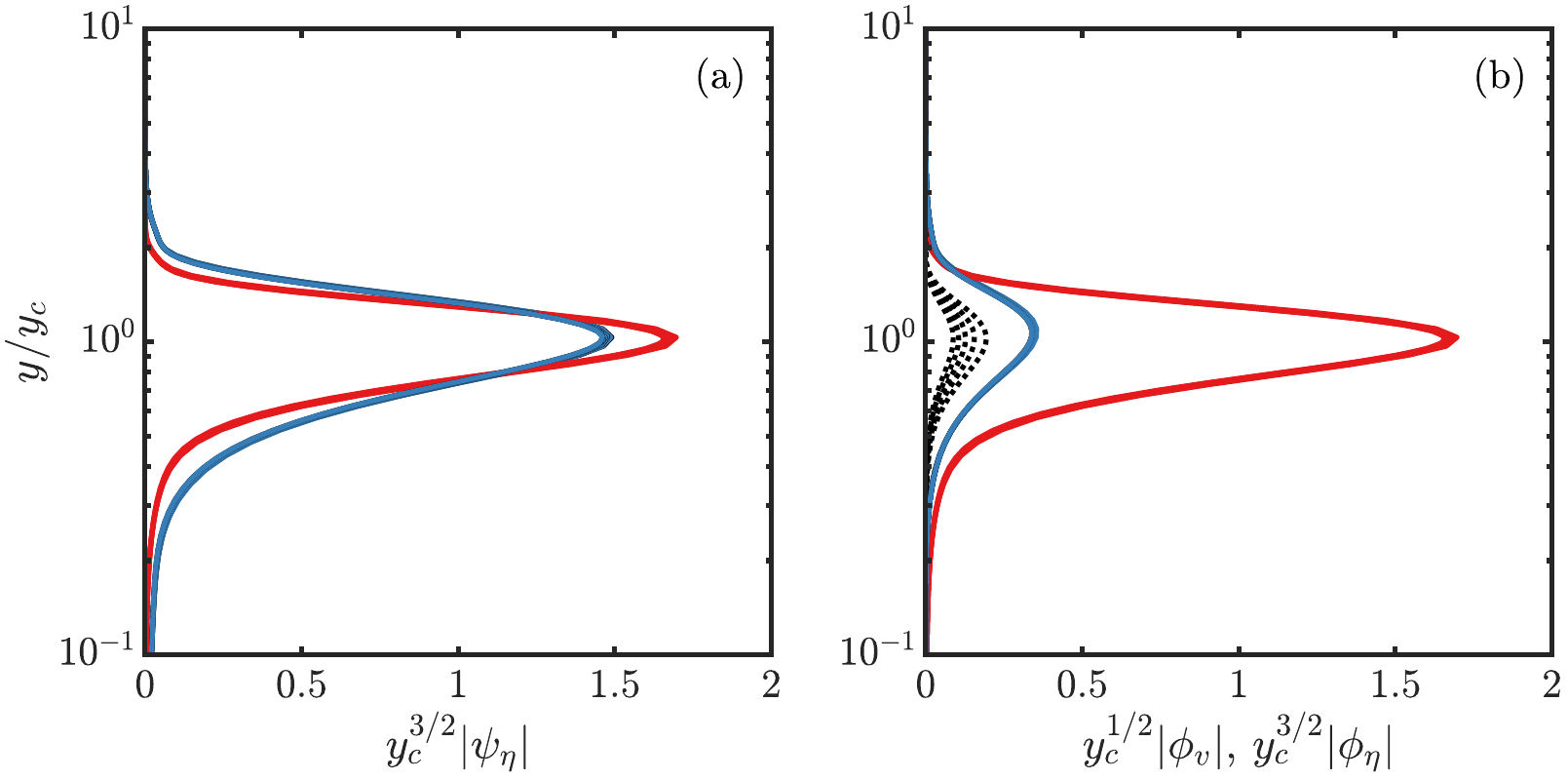}
	\caption{(a) Magnitudes of the scaled OS, SQ, and standard resolvent (color scheme as in~\cref{fig:modes}) self-similar leading vorticity response modes for five members of a hierarchy at $\Rey=2003$, with $k_{x,l} = 10$ and  $k_{z,l} = 10^{3/2}$. (b) Corresponding scaled leading forcing mode magnitudes. $\psi_\eta$ and $\phi_v$ for the standard resolvent (black) are indistinguishable from the OS modes (blue). \label{fig:modesss}}
\end{figure}
The $\yc$-scaled versions of the weighted resolvent operators are
\begin{subequations}
	\label{eq:Hopsss}
	\begin{align}
	\begin{pmatrix}
	\op{F}_v \res_{vv} \op{F}_v^\inv \\
	\op{F}_\eta \res_{\eta v} \op{F}_v^\inv
	\end{pmatrix} 
	&=  
	\begin{pmatrix}
	\yc\ycp \ssim{\op{F}}_v \ssim{\res}_{vv} \ssim{\op{F}}_v^\inv \\
	\yc\yc^{+\,2} \ssim{\op{F}}_\eta \ssim{\res}_{\eta v} \ssim{\op{F}}_v^\inv
	\end{pmatrix}
	\label{eq:Hosss}
	\\
	\op{F}_\eta \res_{\eta\eta} \op{F}_\eta^\inv &= \yc\ycp \ssim{\op{F}}_\eta \ssim{\res}_{\eta\eta} \ssim{\op{F}}_\eta^\inv,
	\label{eq:Hsqss}
	\end{align}
\end{subequations}
so that their leading singular values scale as
\begin{equation}
\sigma_j^\OS = \yc\yc^{+\,2} \ssim{\sigma}_j^\OS, \quad \sigma_j^\SQ = \yc\ycp \ssim{\sigma}_j^\SQ.
\label{eq:sigss}
\end{equation}
As with the outer class modes, the scaling of the OS singular values are only expected to hold for the first several modes. 

{The magnitudes of the leading vorticity response modes of the OS and SQ resolvent operators and their corresponding forcing modes for five members of a particular hierarchy at $\Rey=2003$ are shown in \cref{fig:modesss}(a) and \cref{fig:modesss}(b), respectively, using the scalings derived in \crefrange{eq:uss}{eq:sigss}. For comparison, the leading response and forcing mode for the standard resolvent operator are also shown; again, $\psi_\eta$ and $\phi_v$ are indistinguishable from the OS modes. Here again, $\phi_\eta$ for the standard resolvent (black dotted line in \cref{fig:modesss}(b)) and $\phi_\eta^\SQ$ exhibit different scalings, with the standard resolvent $\phi_\eta = O(\yc^{-3/2}\yc^{+\,-1})$.}

Substituting \cref{eq:uss,eq:sigss} into \cref{eq:A}, we obtain the scaled energy density matrices:
\begin{subequations}
	\begin{align}
	&\begin{array}{c c c}
	\tenscomp{A}^{\OS/\OS}_{uu,ij} = \yc\yc^{+\,4} \ssim{\tenscomp{A}}^{\OS/\OS}_{uu,ij}, &
	\tenscomp{A}^{\OS/\SQ}_{uu,ij} = \yc\yc^{+\,3} \ssim{\tenscomp{A}}^{\OS/\SQ}_{uu,ij}, &
	\tenscomp{A}^{\SQ/\SQ}_{uu,ij} = \yc\yc^{+\,2} \ssim{\tenscomp{A}}^{\SQ/\SQ}_{uu,ij},
	\end{array}
	\\
	&\begin{array}{c}
	\tenscomp{A}^{\OS/\OS}_{vv,ij} = \yc\yc^{+\,2} \ssim{\tenscomp{A}}^{\OS/\OS}_{vv,ij},
	\end{array}
	\\
	&\begin{array}{c c c}
	\tenscomp{A}^{\OS/\OS}_{ww,ij} = \yc\yc^{+\,2} \ssim{\tenscomp{A}}^{\OS/\OS}_{ww,ij}, &
	\tenscomp{A}^{\OS/\SQ}_{ww,ij} = \yc\ycp \ssim{\tenscomp{A}}^{\OS/\SQ}_{ww,ij}, &
	\tenscomp{A}^{\SQ/\SQ}_{ww,ij} = \yc \ssim{\tenscomp{A}}^{\SQ/\SQ}_{ww,ij},
	\end{array}
	\\
	&\begin{array}{c c}
	\tenscomp{A}^{\OS/\OS}_{uv,ij} = \yc\yc^{+\,3} \ssim{\tenscomp{A}}^{\OS/\OS}_{uv,ij}, &
	\tenscomp{A}^{\OS/\SQ}_{uv,ij} = \yc\yc^{+\,2} \ssim{\tenscomp{A}}^{\OS/\SQ}_{uv,ij},
	\end{array}
	\end{align}
\end{subequations}
and the streamwise energy spectrum is 
\begin{equation}
E_{uu} = \yc\yc^{+\,4} \,\Re\!\left\{ \tr\! \left( \ssim{\tens{A}}^{\OS/\OS}_{uu}\tens{X}^{\OS/\OS} \right) \right\} + 2\yc\yc^{+\,3} \,\Re\!\left\{ \tr\! \left( \ssim{\tens{A}}^{\SQ/\OS}_{uu}\tens{X}^{\OS/\SQ} \right) \right\} + \yc\yc^{+\,2} \,\Re\!\left\{ \tr\! \left( \ssim{\tens{A}}^{\SQ/\SQ}_{uu}\tens{X}^{\SQ/\SQ} \right) \right\},
\label{eq:Euuss} 
\end{equation}
where the $\yc$ dependence of the right-hand side is explicit, except for the unscaled weights matrices $\tens{X}^{\text{X}/\text{Y}}$. Balancing all three terms requires 
\begin{equation}
\left\vert \frac{\chi_j^\SQ}{\chi_j^\OS} \right\vert \sim \ycp.
\label{eq:ratss}
\end{equation} 
The ratio $\sqrt{\tenscomp{X}_{11}^{\SQ/\SQ}/\tenscomp{X}_{11}^{\OS/\OS}}$ is plotted along several hierarchies with different $k_{x,l},k_{z,l}$ for $\Rey=2003$ in \cref{fig:wss}. {In this case, to have a sufficient number of wavespeeds belonging to $\mathscr{S}_h$ while keeping the size of the optimization problem manageable, the matching of the DNS spectra is only enforced for $y^+_{\min}=100\leq y^+\leq 0.1\Rey=y^+_{\max}$, with $N_c=25$.} The scaling given by \cref{eq:ratss} is clearly demonstrated, as the data in all cases exhibit a linear dependence on $\ycp$ to within a good approximation. In all cases shown the aspect ratio at the bottom of the hierarchies is $\gamma=5$. Similar results are obtained using different aspect ratios, provided that $\gamma \gtrsim \sqrt{10}$ \cite{moarref2013}. {The slopes of the lines are observed to decrease with increasing $k_{x,l}$. Although not shown here, the slopes tend to increase with increasing $\gamma$.}


\section{\label{sec:conclusions}Discussion and Conclusions}
A low-order {representation of} the time-averaged energy spectra of turbulent channel flow based on the resolvent analysis framework was presented. The resolvent mode weights, which encode information about the nonlinear interactions in the flow were determined empirically by computing the weights that minimize the deviation between the resolvent spectra and spectra obtained from DNS using a convex optimization scheme. The present approach is a modification of previous work \cite{moarref2014}, with the major difference being the incorporation of a recently-proposed alternative decomposition of the resolvent operator into two distinct families of modes, referred to as the Orr-Sommerfeld and Squire families \cite{rosenberg2019a}.

It was demonstrated that the alternative OS-SQ decomposition results in a dramatic improvement in the performance of the representation. This improvement is attributed to the isolation of the $v$ response in the OS family, which enables the $\eta$ response of the SQ family to compete with the large $\eta$ response generated by the OS modes. Furthermore, for certain values of wave the parameters, the leading modes of the standard resolvent operator are almost identical to the leading modes of the OS resolvent, so that the mechanisms encoded in the SQ operator are essentially neglected; this helps explain the relatively poor performance of the representation obtained using the standard resolvent. A decomposition of the statistics into contributions from the OS modes, SQ modes, and an interaction between the two families supports this claim and is in agreement with results from DNS at $\Rey=185$ \cite{rosenberg2018}. It was further shown that the competition between the OS and SQ modes can be interpreted as a phase difference, and that this phase difference is speculated to be close to $\pi$ over large portions of spectral space.

Next, the scaling of the leading singular values for the OS and SQ families were derived for the inner, outer, and geometrically self-similar universal classes of resolvent modes \cite{moarref2013}. For the inner class, both sets of singular values scale as $\Rey^\inv$. For the outer and self-similar classes, the OS singular values are larger than the SQ ones by a factor of $\Rey$ and $\ycp$, respectively. Interestingly, this {large difference in amplification} suggests that modes in these classes are likely to be among those for which the OS and standard resolvent modes are nearly identical. Indeed, the scaling of the leading OS singular values is the same as that for the leading singular values of the standard resolvent for the outer and self-similar classes. For the inner class, the scaling for both the OS and SQ singular values match the standard resolvent. Combining the scalings with the hypothesis that the competition between SQ and OS modes discussed above remains relevant for arbitrary $\Rey$ and throughout the flow domain was used to derive the relative scalings of the OS and SQ weights in each of the universal classes. The scaling predictions were tested against the optimized weights, and, with the exception of high aspect ratio modes of the inner class localized very near the wall, good agreement with the computed optimal weights was found for each of the universal classes.

{The results presented herein have several important implications for equation-driven modeling of turbulent channel flow. The first is that partitioning the resolvent operator into Orr-Sommerfeld and Squire subsystems, originally presented in the context of ECS \cite{rosenberg2019a}, is also advantageous in terms of its ability to develop compact representations of fully turbulent channel flow at high Reynolds number. Furthermore, it provides valuable insight into the complex dynamics by identifying the competition mechanism between the OS and SQ modes, which has ramifications for modeling nonlinear interactions. Specifically,} considering that for large $\Rey$, the OS singular values in the logarithmic and outer regions of the flow are much larger than the SQ ones, it may be tempting from a modeling perspective to neglect the SQ family of modes. However, doing so does not take into account the relative scaling of the forcing terms $\hat{g}_v$ and $\hat{g}_\eta$ in \cref{eq:NSEOS} -- it implicitly assumes they remain of the same order. The present results indicate that this is not the case. In fact, the scaling results of the weights for all of the classes can be summarized as $\vert \chi^\SQ_j/\chi^\OS_j \vert \sim \sigma^\OS_j/\sigma^\SQ_j$. 

Though the absolute scalings of the weights were not determined, the present work can be considered a starting point to guide further modeling efforts toward quantifying nonlinear interactions in turbulent channel flow. {For instance, it is particularly intriguing that, as discussed in \cref{sec:kxspectra}, the $v$ statistics depend only on the OS modes. Consequently, if the scaling of the OS weights can be determined from these, empirically or otherwise, then the results given in \cref{sec:scaling} can be used to determine the scaling of the SQ weights, effectively reducing the number of unknowns by half. Then a single computation at a relatively low Reynolds number could be combined with the scalings to make predictions of the spectra at Reynolds numbers that are currently unattainable by DNS.}

Taken together, the results point to the competition between the OS and SQ modes being an important mechanism in turbulent channel flow that should be respected in order to accurately model the statistics. We hypothesize that if this mechanism could be interrupted, the dynamics, and consequently the statistics, of the system would be significantly different. This line of inquiry is the subject of ongoing work.


\begin{acknowledgments}
The support of AFOSR under grant FA 9550-16-1-0361 and ONR under N00014-17-1-3022 is gratefully acknowledged. Additionally, the authors would like to thank Javier Jim\'{e}nez for making the spectra for the $\Rey=934$ and $\Rey=2003$ simulations publicly available, as well as Adri\'{a}n Lozano-Dur\'{a}n for sharing the spectra for $\Rey=4219$. 
\end{acknowledgments}

\appendix
\section{\label{sec:err}Behavior at large ${\lambda}_x$}
\begin{figure}
	\centering \includegraphics[width=0.46\textwidth]{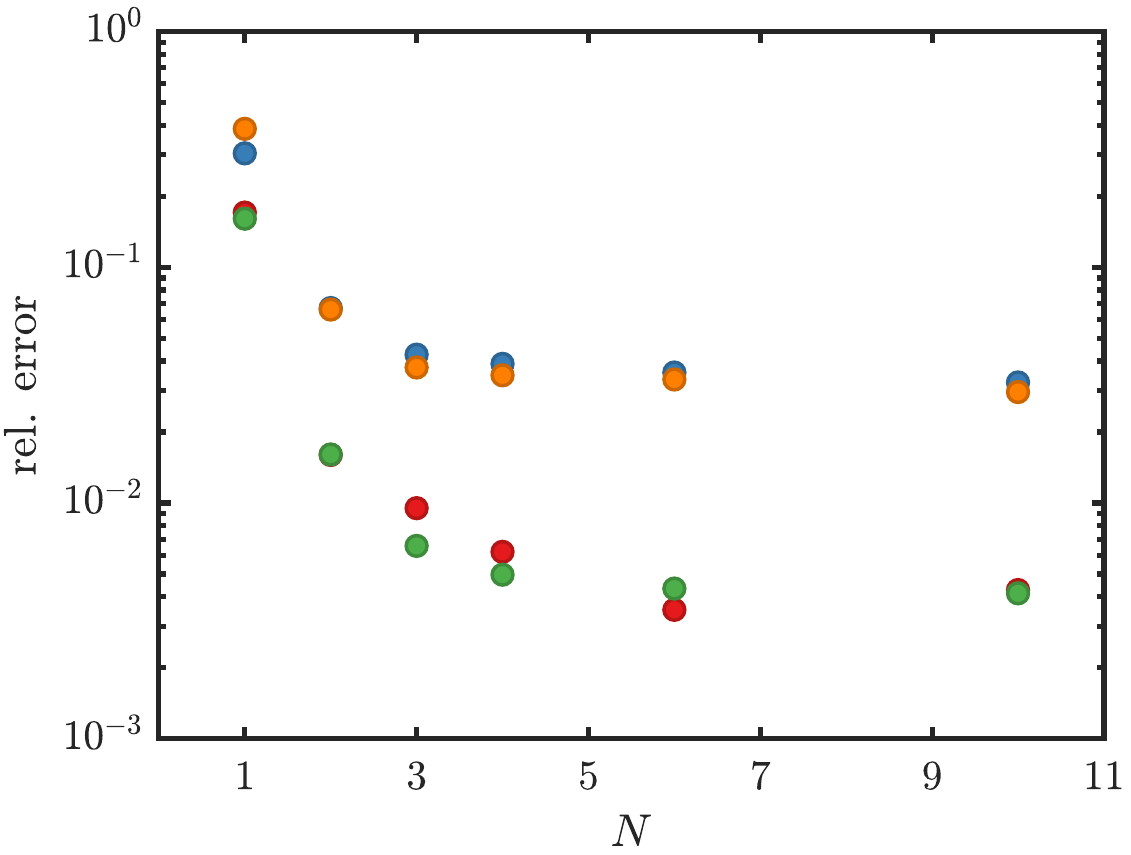}
	\caption{Relative errors in $\langle u^2 \rangle$ (blue), $\langle v^2 \rangle$ (red), $\langle w^2 \rangle$ (green), and $\langle -uv \rangle$ (orange), as a function of $N^{\OS}=N^{\SQ}=N$.\label{fig:intserr}}
\end{figure}

As discussed in \cref{sec:kxspectra}, the most significant discrepancies between the DNS and OS-SQ representation of 1D the spectra shown in \cref{fig:kxspect} occur in $k_xE_{uu}$ at large $\lambda_x^+$ and $y^+ \lesssim 100$ and $-k_xE_{uv}$ at large $\lambda_x^+$ and $y^+ \lesssim 50$. Furthermore, these errors do not improve considerably with an increasing number of modes, as demonstrated in \cref{fig:intserr}, which shows a slow decrease in the error for $N^{\OS}=N^{\SQ}=N > 4$.
\begin{figure}
	\centering \includegraphics[width=0.42\textwidth]{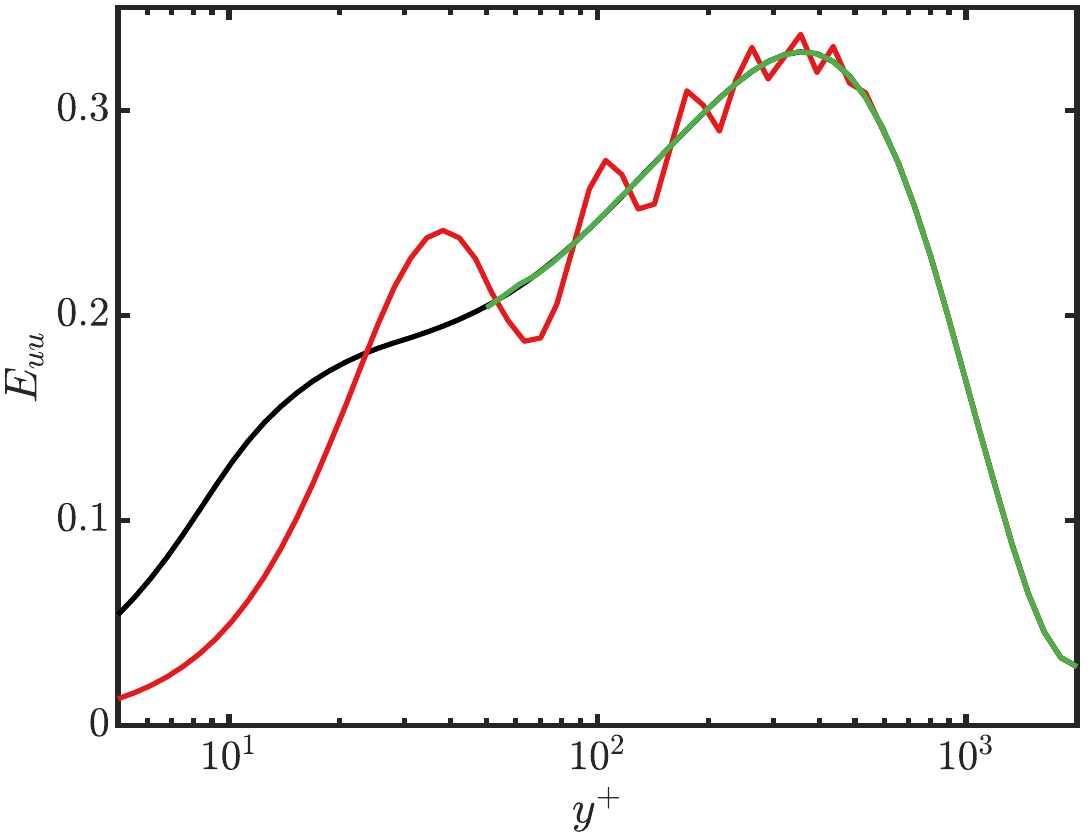}
	\caption{Comparison of DNS (black) and the optimization results with $y^+_{\text{min}}=5$ (red) and $y^+_{\text{min}}=50$ (green) using $N^{\OS}=N^{\SQ}=6$ modes for $(\lambda_x^+, \lambda_z^+) \approx (3.83\times10^4, 2.78\times10^3)$. \label{fig:ypi}}
\end{figure}
The reason for the persisting error is that for large $\lambda_x^+$, there is significant energetic content below the peaks of the {lowest wavespeed 
modes, which typically sit around $y^+\approx 40-50$ for $\Rey=2003$ \cite{moarref2013a}. Thus, trying to match near the wall results in overcompensation at larger $y^+$. This is illustrated for the representative wavelenghts $(\lambda_x^+, \lambda_z^+) \approx (3.83\times10^4, 2.78\times10^3)$ in \cref{fig:ypi}.
To confirm that this is indeed the cause, \cref{fig:ypi} also shows the result of the optimization with $y^+_{\text{min}}=50$, in which case the large oscillations disappear.} 

{The near-wall errors eventually diminish as the number of modes tends to infinity.} However, the fact that the spectra for these wavenumbers are not well-represented by a low-rank approximation suggests that the response modes may not be the most efficient basis. Indeed, \citet{rosenberg2019b} outline conditions under which the flowfields around a cylinder and for channel ECS are more compactly represented by the response to the forcing generated by the leading response at a different wavenumber triplet; it is possible that the present case is a similar situation, but the number of triadic interactions that would have to be accounted for in fully turbulent flow significantly complicates matters. 
It has also recently been shown that augmentation of \cref{eq:resop} with an eddy viscosity improves the representation of large-scale structures 
\cite{hwang2016,illingworth2018,madhusudanan2019}. 
{Both approaches attempt to constrain the forcing, the former by using triadic interactions to identify which scales are most important, and the latter by choosing to only directly model the large-scale coherent motions. However, as pointed out in \cref{sec:intro}, such nonlinear interactions are incompatible with the turbulent mean velocity profile when an eddy viscosity is included.}

\bibliography{references}

\end{document}